\documentclass[preprint]{aastex}

\usepackage{natbib}
\usepackage{longtable}
\usepackage{lscape}
\slugcomment{draft version  \today}
\begin{document}
\title{Properties of the Distant Kuiper Belt: Results from the Palomar Distant Solar System Survey }
\author{Megan E. Schwamb\altaffilmark{1},Michael E. Brown\altaffilmark{1},David L. Rabinowitz\altaffilmark{2}, and Darin Ragozzine\altaffilmark{3}}
\altaffiltext{1}{Division of Geological and Planetary Sciences, California Institute
of Technology, Pasadena, CA 91125}
\altaffiltext{2}{Department of Physics, Yale University, P.O. Box 208121, New Haven, CT 06520}
\altaffiltext{3}{Harvard-Smithsonian Center for Astrophysics, Cambridge, MA 02138}
\email{mschwamb@gps.caltech.edu}

\begin{abstract}
We present the results of a wide-field survey using the 1.2-m Samuel Oschin Telescope at Palomar Observatory. This survey was designed to find the most distant members of the Kuiper belt and beyond. We searched $\sim$12,000 deg$^2$ down to a mean limiting magnitude of 21.3 in R. A total number of 52 KBOs and Centaurs have been detected, 25 of which were discovered in this survey. Except for the re-detection of Sedna, no additional Sedna-like bodies with perihelia greater than 45 AU were detected despite sensitivity out to distances of 1000 AU. We discuss the implications for a distant Sedna-like population beyond the Kuiper belt, focusing on the constraints we can place on the embedded stellar cluster environment the early Sun may be have been born in, where the location and distribution of Sedna-like orbits sculpted by multiple stellar encounters is indicative of the birth cluster size. We also report our observed latitude distribution and implications for the size of the  plutino population. 

\end{abstract}
\section{Introduction}. 

With the advent of wide-field CCD cameras in the past decade, there has been an explosion in observational programs searching for Kuiper belt objects (KBOs) \citep{1995AJ....109.1867J,1996AJ....112.1225J,1998AJ....115.2125J,2000AJ....120.2687S,2001AJ....121..562L,2001AJ....122..457T,2003EM&P...92...99T,2005AJ....129.1117E,2007AJ....133.1247L,2008ssbn.book..335B,2009AJ....137.4917K}. Now there are over 1000 KBOs known, with about half having secure orbits. The majority of these surveys search for distant solar system bodies using images taken on a single night over a span of a few hours, probing out to distances of $\sim$100 AU. Most of these surveys have focused on observing within 10$^\circ$ of the ecliptic with the majority only imaging within just a few degrees.  

The discovery of Sedna \citep{2004ApJ...617..645B} on a highly eccentric orbit far outside the Kuiper belt challenges our understanding of the solar system. With a perihelion of 76 AU, Sedna is well beyond the reach of the gas-giants and could not be scattered into its highly eccentric orbit from interactions with Neptune alone \citep{2003MNRAS.338..443E,2005CeMDA..91..109G}. Sedna's aphelion at ~1000 AU is too far from the edge of the solar system to feel the perturbing effects of passing stars or galactic tides in the present-day solar neighborhood \citep{1987AJ.....94.1330D,1997Icar..129..106F}. Sedna is dynamically distinct from the rest of the Kuiper belt, and its unexpected discovery alludes to a population of icy bodies residing past the Kuiper belt with perihelia greater than 45 AU and semimajor axes greater than $\sim$200 AU, beyond which Neptune is unable to raise the perihelia of scattered disk KBOs through resonant perturbations \citep{2005CeMDA..91..109G}. 

Sedna is the only body known to reside in this region. Sedna was found near perihelion at a distance of $\sim$88 AU, at the motion limit and brightness limit of its discovery survey (Brown et al 2008). With one night imaging, previous KBOs surveys were likely insensitive to the objects in the Sedna region. To date, surveys \citep{2007AJ....133.1247L,2008ssbn.book..335B,2010arXiv1004.3288P} have been unsuccessful in finding additional Sedna-like bodies. In order to find the largest and brightest members of the Sedna population, we have been engaged in an observational campaign to survey the northern sky. We present the results of our search for distant solar system bodies covering $\sim$12,000 deg$^2$ within 30$^\circ$ of the ecliptic. Rather than searching over a single night, we use a two-night baseline to distinguish the extremely slow motions of these distant bodies from background stars. We are sensitive to motions out to a distance of  $\sim$1000 AU ($\sim$0.2 $^{\prime\prime}$hr$^{-1}$). 

In this paper, we discuss the implications for a distant Sedna-like population beyond the Kuiper belt and provide constraints on the cluster birth Sedna formation scenario \citep{2006Icar..184...59B}. The survey was specifically designed to find the select brightest members of a distant Sedna population but was also sensitive to the dynamically excited off ecliptic populations of the Kuiper belt including the hot classicals, resonant, scattered disk, and detached Kuiper belt populations. We present our observed latitude distribution and implications for the plutino population. 

\section{Observations}

\begin{figure}
\epsscale{.5}
 \plotone{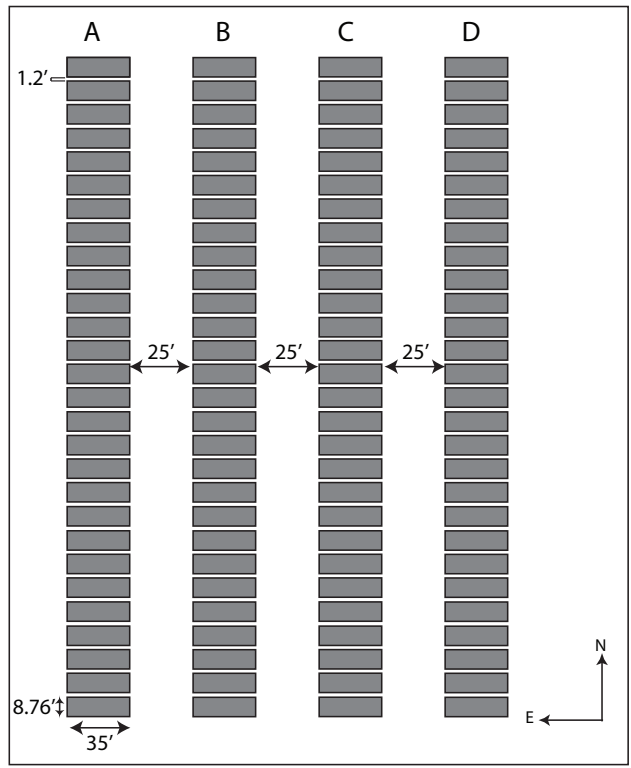}
\caption{Scale drawing of the focal plane of the QUEST camera, depicting  the layout of the 112 CCDs. The gap between CCDs in the north-south direction is $\sim$1.2$^\prime$ and the spacing between adjacent fingers along the east-west direction is  $\sim$25$^\prime$. }
\label{fig:quest}
\end{figure}

Observations were taken nightly using the robotic 1.2 m Samuel Oschin Telescope located at Palomar Observatory and the QUEST large-area CCD camera. The QUEST camera has an effective field of view of 8.3 deg$^2$ with a pixel scale of 0.87$^{\prime\prime}$ \citep{2007PASP..119.1278B}. The 161-megapixel camera is arranged in four columns or ``fingers'' along the east-west direction each equipped with 28 2400x600 CCDs in the north-south direction (see Figure \ref{fig:quest}). The gap between CCDs in the north-south direction is $\sim$1.2$^\prime$ and the spacing between adjacent fingers along the east-west direction is  $\sim$25$^\prime$. The four fingers are labeled (A-D) and the CCDs are numbered sequentially (1-28) from North to South. We will refer to the CCDs by finger and position along the finger (i.e.,C14, D28). 

\begin{figure}
\epsscale{1}
\plotone{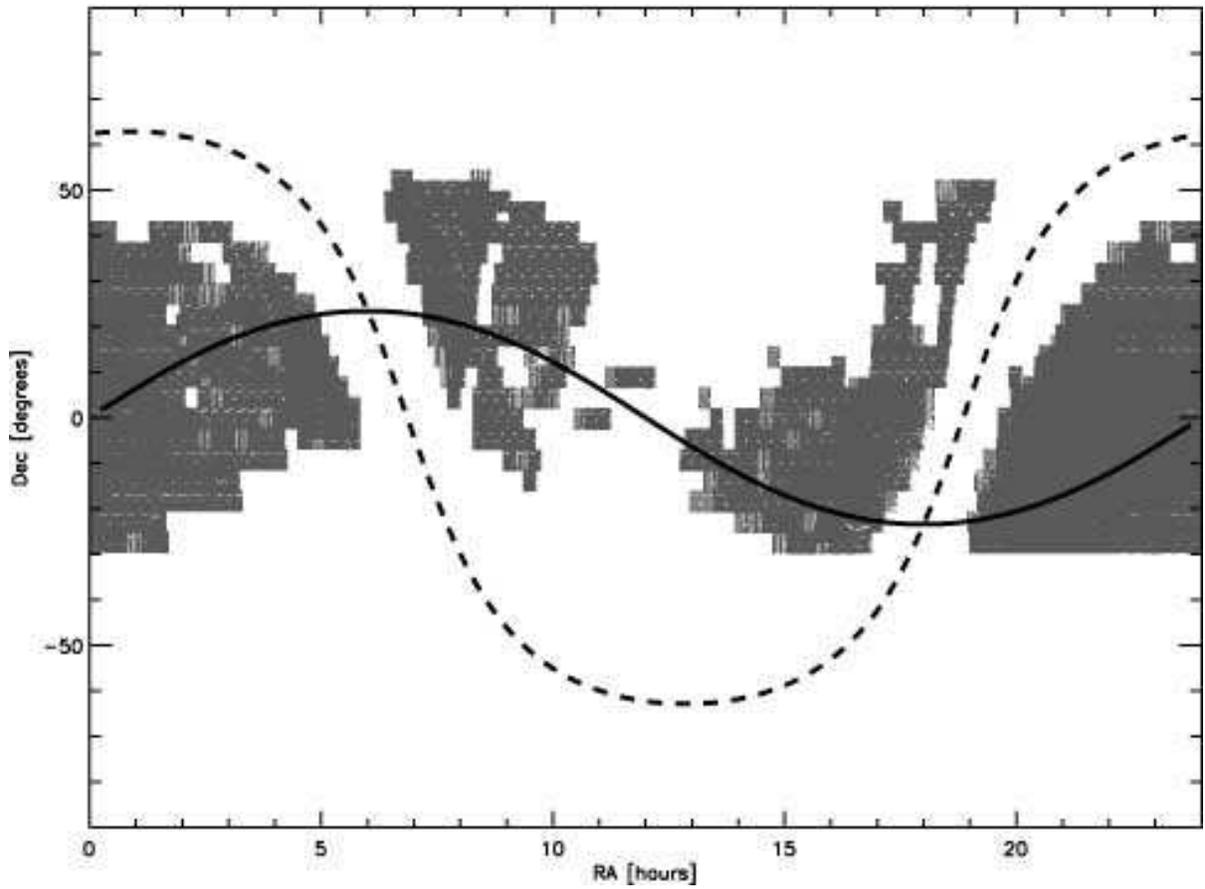}
\caption{Sky coverage of the Palomar survey plotted on the J2000 sky. The observed fields are plotted to scale. The plane of the Milky Way is denoted as a dashed line, and the ecliptic is denoted as a solid line. Holes are due to galactic plane avoidance, bad weather, forest fires, and hardware malfunctions.}
\label{fig:sky}
\end{figure}

Observations were taken from 2007 May 8 - 2008 September 27. We have surveyed in total 11,786 deg$^2$ within $\pm$30$^\circ$ of the ecliptic to a mean depth of R magnitude 21.3. 
Our sky coverage is shown in Figure \ref{fig:sky}. Field centers are compiled in Table \ref{tab:B15} \footnote{\texttt{the full version of Table 1 will be available in the online journal version}}.  A forest fire on Palomar Mountain prevented observations in 2007 September and camera malfunctions ceased operations from 2008 February -2008 May leading to gaps in longitudinal coverage. After 2008 May normal observations resumed until the QUEST camera ceased operations on the Oschin telescope at the end of 2008 September. 

Target fields were observed over a two-night baseline in order to search for solar system objects out to distances of  $\sim$1000 AU (moving at speeds as low as 0.2 $^{\prime\prime}$hr$^{-1}$). All exposures were taken through the broadband red RG610 filter (IIIaF filter from the POSS-II survey) with a wavelength range of $\lambda$=610$-$690nm \citep{1991PASP..103..661R}. For each field, a pair of 240s exposures was taken separated by  $\sim$1 hour on each of the two nights. The second night of observations was typically the next day or at most four nights later. Observations were in varying photometric conditions and lunations. To check the photometric quality of each nightly pair of observations,  magnitudes of the detected sources from both images were histogram binned  with a bin size of 0.2 mag, and the peak value of the histogram was selected as an indicator of image depth.  If the median value of the five CCDs best CCDs (B11,C19,D09,D12,D13) was less than 20.4 mag  (19.0 mag for crowded fields with  greater than 4000 detected sources) than the observation was rejected as poor quality, and the target field was rescheduled for new observations the next night. If a target field cannot be successfully imaged within four nights of the first pair of observations, the field was reset and scheduled for another two nights of observations. 

All target fields were observed within 42$^\circ$ of opposition, and to avoid high star densities, fields less than 15$^\circ$ from the galactic plane were avoided. The camera RA CCD gap was covered by adjacent pointings, but the  $\sim$1.2$^\prime$ declination gap remains mostly uncovered in our survey observations. When all opposition fields within  $\pm$30$^\circ$ of the ecliptic for a month's lunation were completed, overlap pointings were then targeted to reduce holes in our sky coverage due to the camera's declination gap and defective CCDs. From the beginning of the survey to 2007 November 12, instead of performing overlapping coverage, fields with ecliptic latitudes greater than 30$^\circ$ were instead targeted once all available opposition fields within 30$^\circ$ of the ecliptic were completed. 

\section{Data Analysis and Object Detection}

\subsection{Moving Object Detection}
Observations were processed nightly though an automated reduction pipeline using the Interactive Data Language (IDL) Software package. Each CCD on the detector was reduced and searched for moving objects independently from the other CCDs on the mosaic. All images were bias-subtracted and flat-field corrected. A row-by-row median of the overscan region was used for the bias subtraction. A master flat-field image for each CCD was constructed from a 3-$\sigma$ clipped median of the night's science images. Some of the camera CCDs had a significant fraction of hot or defective pixels. These pixels were identified as those where the flat field image value deviated by more than 0.7$\%$ from the value of the 3x3 median boxcar filtered flat field image. To mask the effects of these hot pixels, those regions of the science image were replaced by the median value of a 3x3 pixel box centered on the bad pixel. If hot/bad pixels constituted more than 20$\%$ of the image than a 3x3 median boxcar smoothed image was substituted for analysis. 

SExtractor \citep{1996A&AS..117..393B} was run on each image to compile a list of sources. SExtractor was tuned such that source detection constituted 4 or more contiguous pixels (DETECT$\_$MINAREA parameter) above the detection threshold (DETECT$\_$THRESH  parameter) of 1.2 $\sigma$ above the sky background. Chips C14, C04, and D26 had significant image defects and higher SExtractor detection thresholds were used for these three CCDs (DETECT$\_$MINAREA $=$5 and  DETECT$\_$THRESH $=$2.3) SExtractor performed circular-aperture photometry using a 5-pixel radius aperture. Each source was characterized by its position, flux, and shape. The best of the four images was selected as the master template whose astrometric solution was found by matching image stars to the USNO A2.0 catalog \citep{1998usno.book.....M}. The other three images were then aligned relative to the stars in the template image. Even if the absolute astrometry failed, the relative astronomy between the images was still sufficient to search for distant solar system bodies. The median absolute astrometric error for the entire survey was 0.4$^{\prime\prime}$. The median relative astrometric error between survey images was 0.076$^{\prime\prime}$. 

Once astrometric solutions had been found, the images were searched for moving objects. Because our observations were taken at or near opposition, slow-moving solar system objects were identified by their retrograde motion due to the parallax caused by the Earth's orbital motion. Distant planetesimals may move too slowly to show apparent motion over the nightly one-hour baselines and appear stationary on individual nights. To ensure the detection of objects out to distances of $\sim$1000 AU, we only required motion to be identified over the two-night baseline. The detection catalogs from all four images were compared to identify and eliminate the stationary sources in each image. Sources on one image that had a counterpart within a 4$^{\prime\prime}$ radius on either of the second night's observations were removed as background stars. To further cull the object lists of stars that were above the SExtractor thresholds on one night but below the detection limit on the other, we generated SExtractor source catalogs with more sensitive detection parameters (DETECT$\_$MINAREA = 3 and DETECT$\_$THRESH = 1.1), and compared these deep catalogs to our detection lists. Image sources from one night that appeared on the other night's deep detection catalogs were deemed stationary and rejected as well. Saturated stars and extended sources whose peak flux was more than 3 pixels from the source center measured by SExtractor were also removed from the object catalogs.

Potential moving candidates were identified from the remaining unmatched sources. The nightly images were searched for moving object pairs with motions less than 14.4 $^{\prime\prime}$hr$^{-1}$, the velocity of bodies at distances of 10 AU.  Moving object pairs from the first night and pairs from the second night separated by more than 4.38$^{\prime\prime}$ with retrograde motion consistent with opposition were linked. To eliminate stationary image sources that had been linked between the two nights, candidates with average nightly magnitudes differing by more than one magnitude were eliminated. Remaining candidates whose nightly motions differ by less than twice the first nightÕs on sky velocity were kept  to create the list of moving object candidates. Candidates were filtered via the orbit-fitting package described in \cite*{2000AJ....120.3323B}. Those candidates with successful orbit fits which produced a $\chi^2$ less than 25 and barycentric distance between 15 and 1000 AU were identified as moving objects and added to the final list of candidates to be screened by eye. 100x100 pixel subimages for each of the final moving object candidates were created from the discovery images. These snapshots were aligned and blinked by eye. A total of 39,110 candidates ($\sim$200 a night) were visually inspected. Typical false positives included diffraction spikes, faint background stars, blended sources, and CCD imperfections. 

\subsection{Recovery Observations}

At discovery, heliocentric distance and inclination can be identified from the parallax effect due to the Earth's motion, but other orbital parameters remain unconstrained. With only a two-night discovery arc, a distant Sedna-like body cannot be distinguished from a typical scattered disk Kuiper belt object near aphelion. Even with follow-up observations a month after discovery, both families of orbits provide reasonable astrometric fits to the observations. The two orbital solutions diverge sufficiently a year after discovery, and a secure dynamical identification can only be made after these additional observations. 

Recovery observations of new discoveries were taken at the Palomar 60-inch telescope, the Palomar 200-inch telescope, the 0.9-m  telescope operated by the SMARTS consortium at Cerro Tololo Inter-American Observatory , the 42-inch John S. Hall Telescope located at Lowell Observatory, the 2.66-m Nordic Optical Telescope located at el Roque de los Muchachos Observatory, and then 8.2-m Subaru Telescope on Mauna Kea. Of our detected KBOs, 96$\%$ have multi-opposition observations. All but two discoveries classified as KBOs by the Minor Planet Center (2007 JF45 and 2007 PS45) were recovered during the survey.  The two unrecovered objects were discovered during reprocessing of the data with more sensitive SExtractor source detection parameters and were discovered after they were no longer observable.  Observations taken near 40$^{\circ}$ from opposition, contained contamination from asteroids near their stationary points that appeared to be moving at rates similar to distant KBOs. Some were identified with subsequent observations that confirmed these objects were on orbits with semimajor axes less than 5 AU.  All other objects not successfully recovered have either been linked with other asteroid observations or have been classified on orbits well short of the Kuiper belt by the Minor Planet Center (MPC) database\footnote{\texttt{ http://www.cfa.harvard.edu/iau/Ephemerides/Distant/index.html}}. 

\subsection{Calibration and Efficiency}

\subsubsection{Limiting Magnitude}
\label{sec:limitingmag}
The survey observations were taken during a wide  variety of photometric, seeing, and weather conditions. Each CCD frame was independently photometrically calibrated. A photometric zero point offset to our instrumental magnitudes was derived relative to the USNO A2.0 catalog \citep{1998usno.book.....M} red magnitude. The photometric uncertainty of the USNO catalog is non-negligible. For magnitudes greater than 17, the uncertainty is 0.3 mag \citep{1998usno.book.....M}. We likely have several tenths of magnitude uncertainty in our discovery magnitudes. We have not precisely calibrated the survey depth with calibration observations. Limiting magnitudes were computed based on the USNO catalog. We found that the faintest magnitude with a 5$\sigma$ (10$\sigma$ for C2 A19, C14, C04, D26; CCDs with larger numbers of hot pixels), uncertainty as reported by SExtractor represented an accurate measure of the source detection limit of our images, and we used these values in the work presented in this paper. The limiting magnitude for each nightly pair of field observations was taken as the depth of the shallower of the two images.The mean limiting magnitude of the survey based on the USNO catalog is 21.3 in R. 

\subsubsection{Survey Efficiency}
\label{sec:se}
Because our survey has covered a wide swath of sky detecting multiple previously known KBOs, we have an alternative method of determining the limiting magnitude of our survey. Of our detections, 27 are previously discovered KBOs and Centaurs in the MPC database. The absolute magnitudes recorded in the MPC are based upon the apparent magnitudes measured from the discovery or follow-up observations, like our survey, which are often taken in non-standard filters and observed without precise photometric calibrations. \cite{2005Icar..179..523R} find the absolute magnitudes recorded in the MPC are systematically 0.3 mag brighter than those magnitudes accurately measured for their sample of 90 KBOs and Centaurs. We can still use the known population of bright KBOs to estimate a crude efficiency for the survey.  We obtained the positions and visual apparent magnitudes computed by JPL Horizons\footnote{\texttt{http://ssd.jpl.nasa.gov/horizons.cgi}} for known KBOs. As of 2010 January 20, there were 64 previously known multiopposition KBOs with visual magnitudes brighter than 22nd magnitude (excluding discoveries found in this survey and objects with $a$ $<$ 30) with predicted positions located on our survey images that could have been detected by our detection pipeline. We only considered KBOs positioned on the same CCD for all 4 field observations, not accounting for masked regions of the CCDs. Masked bad pixel regions account for $\sim$8$\%$ of the QUEST camera's observable area, but  likely the loss due to bad pixels is much smaller than this value. A KBO positioned on a bad pixel may not necessarily be lost, Sextractor interpolates values for masked pixels from neighboring good pixels before source detection. 
 For every object not detected in the survey, we examined the images to determine if a moving source was visible. No known KBO was missed during the visual inspection of moving object candidates. The majority of the missed KBOs were not found because the KBO's psf overlapped with a neighboring star and was missed by SExtractor, the KBO was on a bad or masked off region of the CCD, image quality was bad due to poor telescope tracking, or the KBO was too faint to be detected and no visible moving source was identifiable. 

\begin{figure}
\plotone{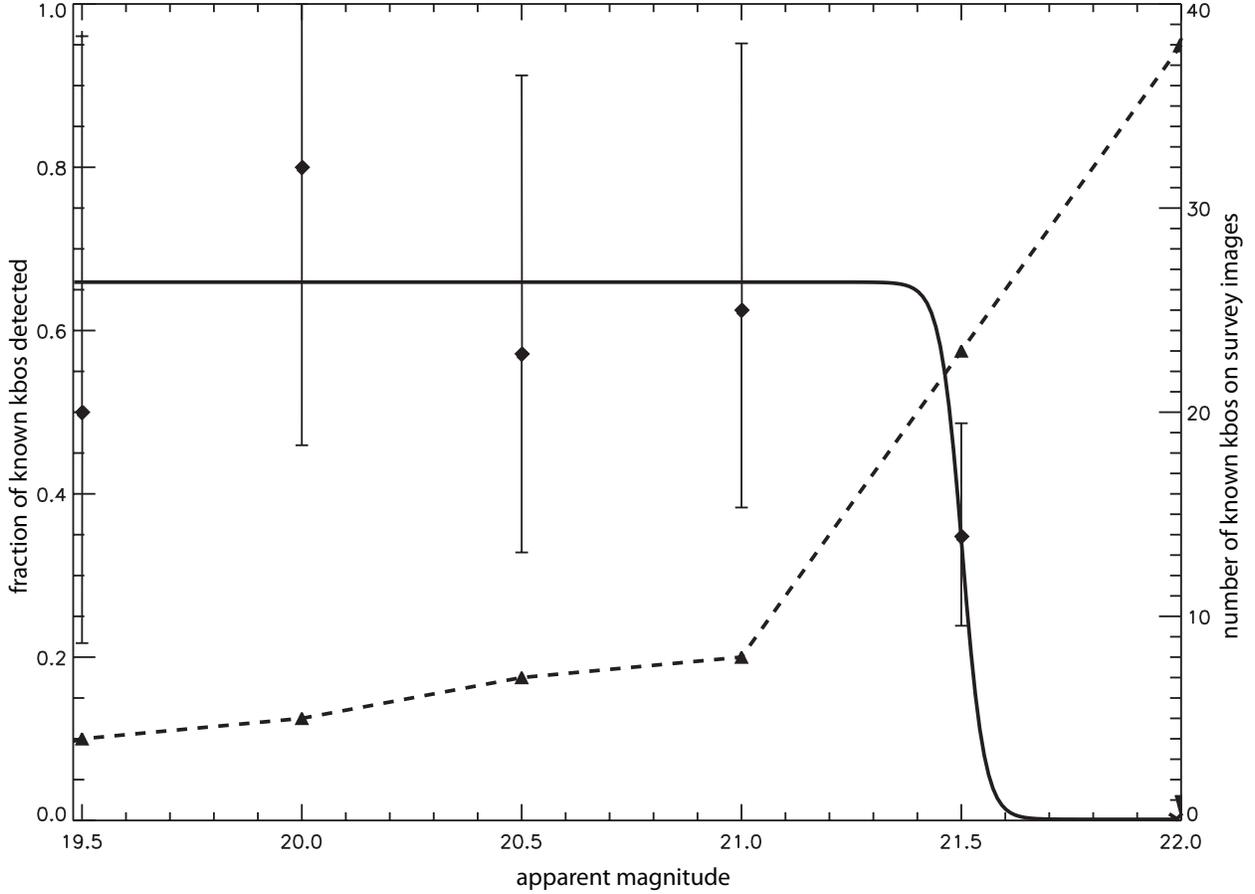}
\caption{Survey efficiency based on the previously known multiopposition KBO population.Solid line plots the  best-fit efficiency function. Diamonds plot the binned detection efficiency in 0.5 magnitude bins with one-$\sigma$  Poissonian error bars.  The dashed line with triangles is the number of previously known multiopposition KBOs ($a$ > 30 AU) in each magnitude bin with predicted positions located on all 4 survey images. }
\label{fig:eff}
\end{figure}

We define the survey efficiency function as: 

 \begin{equation}
\varepsilon= \frac{\varepsilon_{max}}{2}  \left( 1-\tanh\left(\frac{m-m^*}{g}\right)\right)
\end{equation}
where $\varepsilon$  is the efficiency with which KBOs of magnitude $m$ are detected in our survey, $\varepsilon_{max}$ is the maximum efficiency, $m^*$ is the magnitude at which $\frac{\varepsilon_{max}}{2}$, and $g$ is the half width. We fit for the efficiency by computing the cumulative distribution for all known KBOs scaling for the probability of detection and compare to the observed cumulative distribution. To find the optimal parameters,  we minimize the $\chi^2$ between the observed and calculated cumulative distributions. We find  $\varepsilon_{max}$=0.66, $m^*$=21.5, $g$=0.05. The efficiency drops by 50$\%$ at 21.5 V mag, consistent with our median image limiting magnitude. Figure $\ref{fig:eff}$ plots the best-fit efficiency function and plots the binned detection efficiency for the known sources located on all 4 field images in 0.5 mag bins. We estimate the uncertainty in our survey efficiency using the number of known KBOs found with magnitudes less than or equal to 21st magnitude, well before the drop off in the best-fit efficiency function. We found 13 of 19 known KBOs brighter than or equal to 21st magnitude, giving an efficiency of 68$\%$, consistent with our best-fit efficiency function, and assuming Poisson counting statistics, the 1-$\sigma$ confidence level ranges from 51-89$\%$. 

\subsubsection{Geometric Losses}

The gap between the QUEST camera's CCDs in the north-south direction is $\sim$1.2$^\prime$. Along the east-west direction the separation between CCDs is  $\sim$25$^\prime$.  At the distances our survey is sensitive to, a KBO located in the declination gap would remain in the gap between CCDs over the two-night baseline. This was the case for Eris, and Eris was not detected in our survey. For some areas of the sky we do have overlap pointings  to try and cover the declination gap but only after all opposition target fields were observed. The losses  due to the CCD gaps is accounted for in our sky coverage estimates, but we do not include the loss due to masked regions. 

\begin{figure}
\plotone{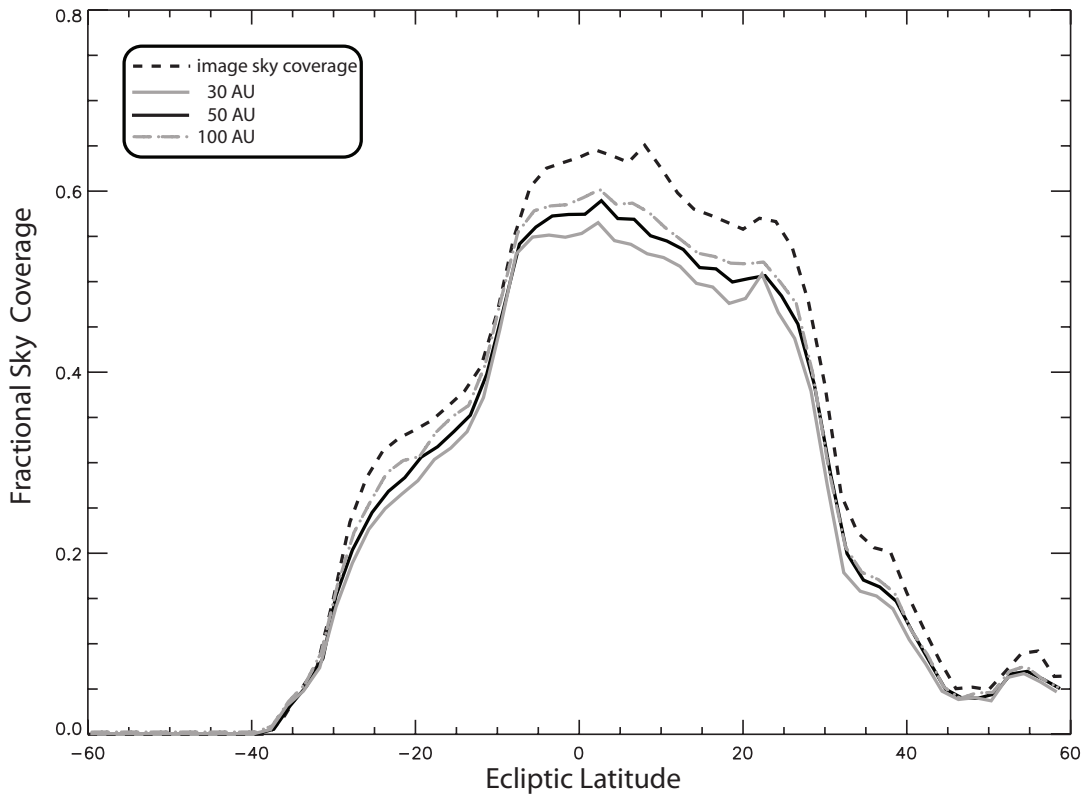}
\caption{Effect of geometric losses on the sky coverage of the survey. The fraction of simulated orbits found on all four CCDs as a function of latitude binned in two degree bins. Main effects are due to KBOs that are not located on all field observations and move off the CCD or objects positioned in the gaps between the CCDs.  }
\label{fig:geometry}
\end{figure}

KBOs that moved off the edge of the CCD into the CCD gaps were missed by our automated detection pipeline. Non-functioning CCDs and longitudinal losses are accounted for in our latitudinal sky coverage estimates, but to measure our geometric losses from those KBOs moving off the CCDs or lost in the CCD gaps,  we generated $\sim$10$^6$ random circular orbits assuming a uniform inclination distribution (0-180$^\circ$) for a range of semimajor axes. Neglecting the effect of masked CCD regions, we calculated the fraction of simulated KBOs positioned on all four survey images as a function of ecliptic latitude.  Figure \ref{fig:geometry} compares our survey sky coverage to the fractional coverage of  the simulated circular orbits at 30, 50, and 100 AU. The greatest losses occur at the ecliptic, and we find this effect is at most $\sim$10$\%$. Closer orbits are moving at faster on-sky velocities and are more likely to move off the CCD over the two night-base line than objects at further distances, but we find the difference in losses by objects at 30 and at 100 AU is small, and that all objects in the Kuiper belt have similar geometric losses in our survey. 

\subsubsection{Pipeline Detection Efficiency of Sedna-like Bodies}

\begin{figure}
\plotone{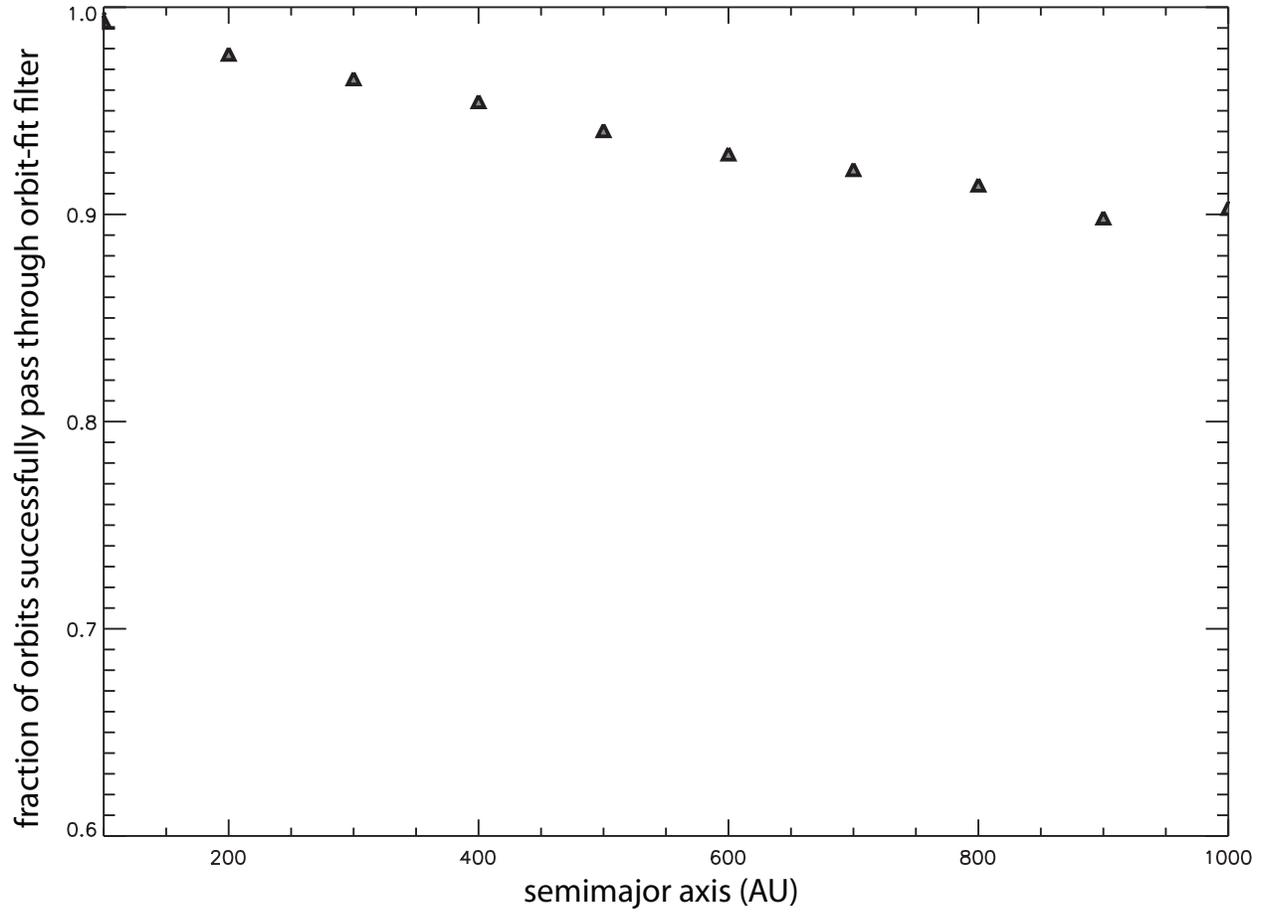}
\caption{Efficiency of our orbit-fit detection filter. Fraction of synthetic orbits with barycentric distances between 15-1000 AU that successfully are identified as outer solar system bodies found on all four survey images versus semimajor axis binned in 100 AU bins  }
\label{fig:sedna_eff}
\end{figure}

Any comparison of the Sedna population requires that we also understand whether these bodies would be detected in our survey. Many of the mechanisms proposed for the formation of Sedna  \citep{2004Natur.432..598K, 2004AJ....128.2564M,2006Icar..184...59B,2007Icar..191..413B,2008Icar..197..221K} produce many highly eccentric and even retrograde orbits.   To test whether Sedna-like orbits would pass through our orbit-fitting filter, we created artificial orbits with a uniform semimajor axis ranging from 100-1100 AU and uniform eccentricity and inclination distribution including retrograde orbits. For those 781,763 artificial orbits whose positions land on our images, have barycentric distances less than 1000 AU and have perihelia greater than 50 AU, we add absolute and relative positional offsets characteristic of the survey's astrometric errors. All four images of a field observation have the same absolute astrometric error but random relative positional errors.  We add normally distributed random absolute and relative astrometric errors using the three-sigma clipped median and standard deviation of the survey astrometric uncertainties. As shown in Fig \ref{fig:sedna_eff}, the efficiency is the fraction of synthetic orbits fit with the \cite*{2000AJ....120.3323B} software that pass our selection criteria in each semimajor axis bin compared to the number of objects in the 100 AU bin. 5$\%$ of the simulated orbits would not have made it through to visual inspection with the majority of failures due to the best-fit orbit placing the object on an asteroid-like orbit. We are confident that Sedna-like bodies present in our images detected by SExtractor would be identified by our automated detection scheme.  

\section{Detections}

\begin{figure}
\epsscale{.75}
\plotone{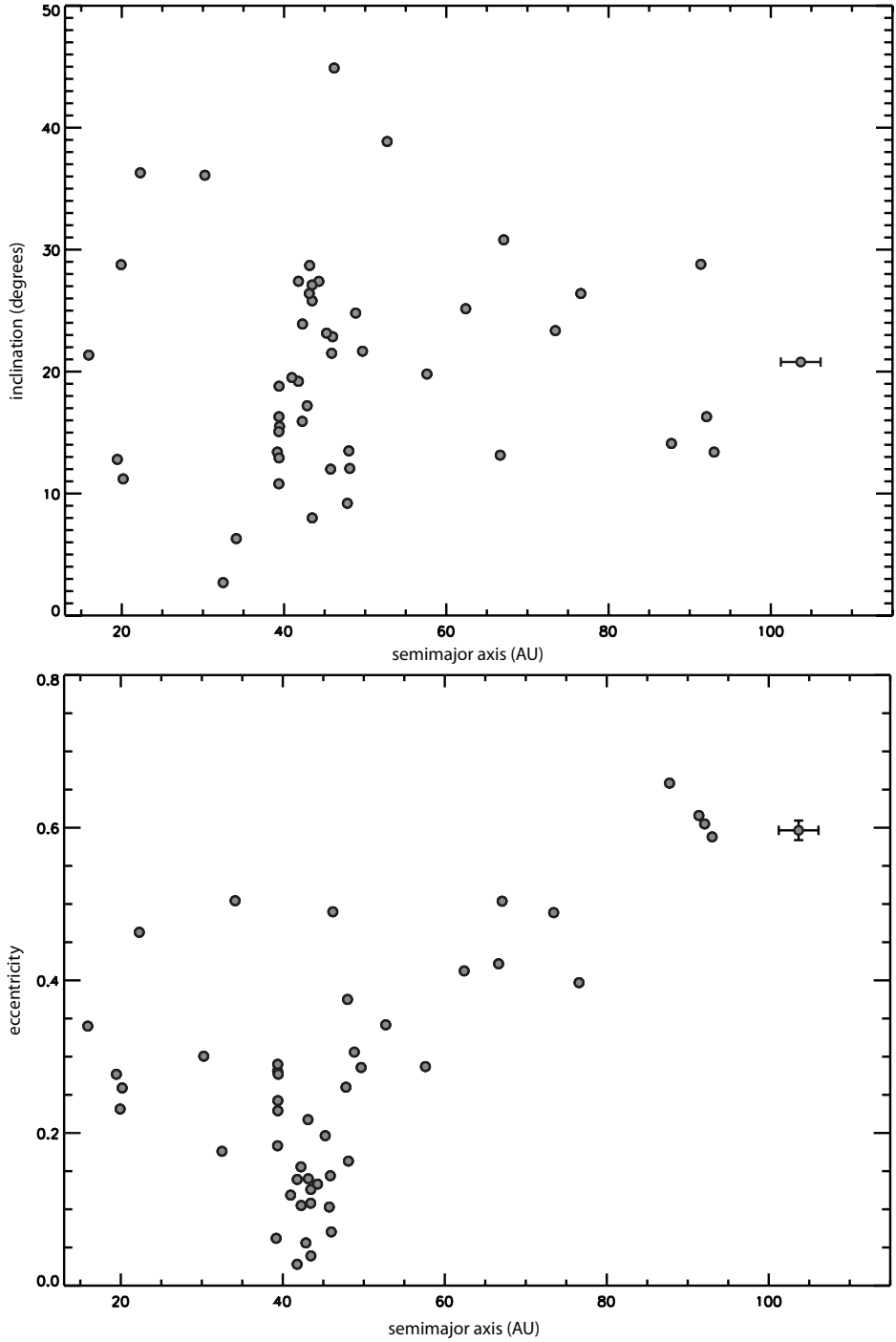}
\caption{Eccentricity vs. Semimajor axis and Inclination vs. Semimajor axis of multiopposition objects found in the Palomar survey. Sedna has been excluded for better resolution. One-$\sigma$ errors from \cite{2000AJ....120.3323B} orbit fit are plotted. The error bars are typically smaller than the size of the symbol. }
\label{fig:orbits}
\end{figure}

\begin{figure}
\epsscale{1}
\plotone{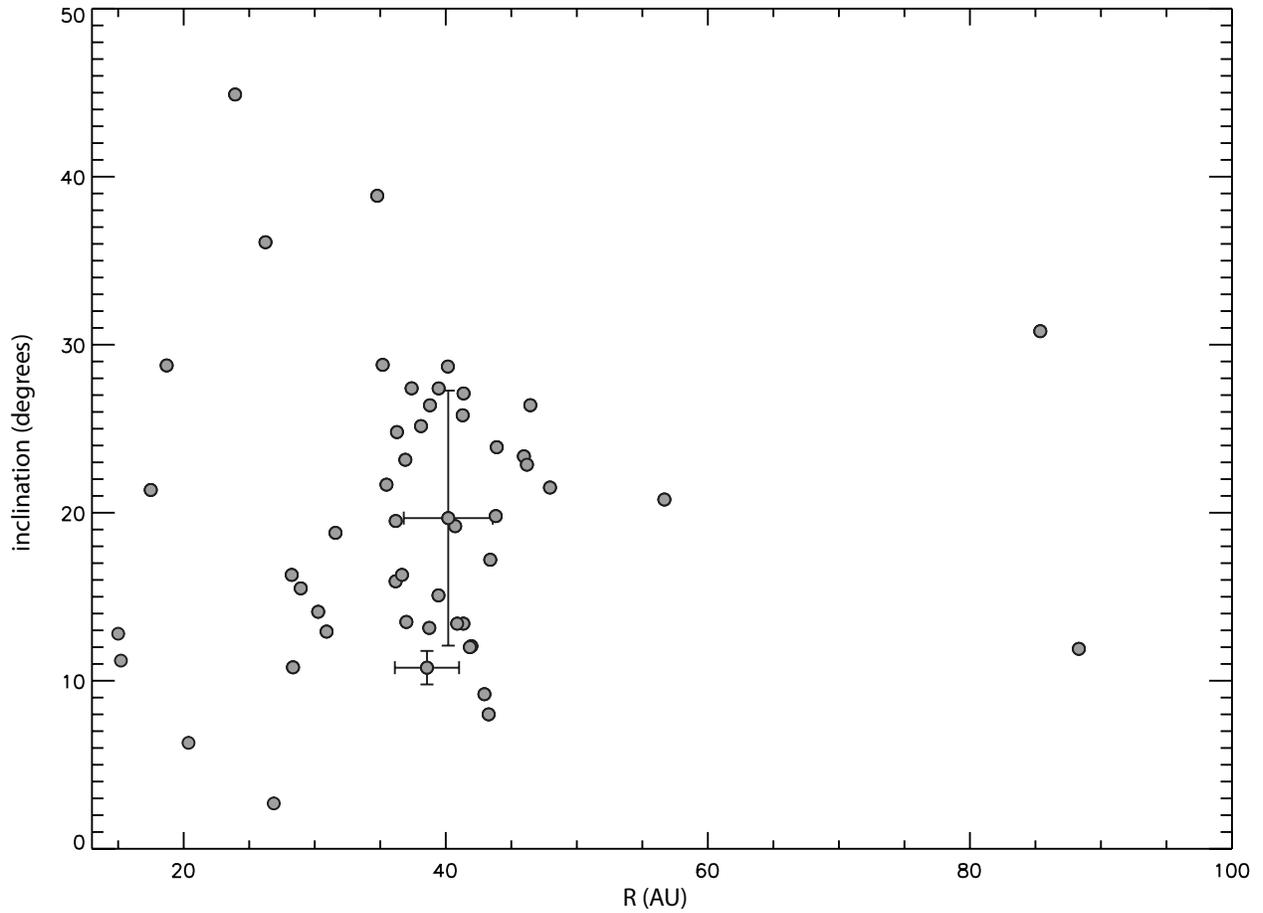}
\caption{Inclination vs. barycentric distance at discovery for objects detected in the Palomar survey. One-$\sigma$ errors from \cite{2000AJ....120.3323B} orbit fit are plotted. The error bars are typically smaller than the size of the symbol. }
\label{fig:radial}
\end{figure}

A total of 52 KBOs and Centaurs have been detected of which 25 are new discoveries from this survey. 50 of our discovered objects have multiopposition orbits. Table \ref{tab:kbos} lists the orbital information for objects detected in the survey. The orbital and radial distribution is plotted in Figure \ref{fig:orbits} and Figure \ref{fig:radial} respectively. With the detection of no cold classical belt objects; our overall survey probes the orbital properties of the hot classical, scattered disk, detached, and resonant populations. The survey was specifically designed to probe the Sedna region, but except for Sedna, no additional objects with perihelion greater than 45 AU were detected despite sensitivity out to distances of 1000 AU.  

We do detect several Centaurs with semimajor axes less than 30 AU in our survey, but to constrain the number of false detections by inner solar system objects we placed a minimum distance threshold in our moving object detection scheme. Candidates with barycentric distances less than 15 AU as calculated from initial orbits fit by the \cite*{2000AJ....120.3323B}  method were ignored. Our survey is limited to detecting only the most distant of the Centaurs, and we therefore will not address the Centaur population in this paper. 

\section{Sedna Population}

With a perihelion of 76 AU and an aphelion of $\sim$1000 AU Sedna is dynamically distinct from the rest of the Kuiper Belt. Its extreme orbit suggests the presence of a population of icy bodies residing past the Kuiper belt. The study of this Sedna population provides a unique new window into the history of the early solar system. Some other mechanism no longer active in the solar system today is required to emplace Sedna on itÕs highly eccentric orbit. Several possible scenarios have been offered to explain Sedna's extreme orbit, including interactions with planet-sized bodies \citep{2006ApJ...643L.135G, 2006Icar..184..589G,2008AJ....135.1161L,2010IAUS..263...67G}, stellar encounters \citep{2004AJ....128.2564M}, multiple stellar fly-bys in a stellar birth cluster  \citep{2004AJ....128.2564M,2006Icar..184...59B,2007Icar..191..413B,2008Icar..197..221K}, interstellar capture \citep{2004Natur.432..598K,2004AJ....128.2564M}, and perturbations from a wide-binary solar companion  \citep{2005EM&P...97..459M}.  Each of the various Sedna formation models leave a distinctive imprint on the members of this class of distant objects and has profound consequences for our understanding of the solar system. These planetesimals in the Sedna region are dynamically frozen and the relics of their formation process. The orbital distribution and number density of Sedna-like bodies will distinguish between the formation scenarios. 

In \cite{2009ApJ...694L..45S}, before recovery observations were complete, we compared the expected number of detections from a theoretical population on orbits with the same semimajor axis and eccentricity as Sedna to our survey results, the redetection of Sedna. Our best-fit value gives 40 bodies residing on Sedna's orbit that are brighter than or equal to Sedna. At the one-$\sigma$ confidence level we ruled out a population larger than 92 and smaller than 15 Sedna-sized or bigger objects on orbitÕs similar to Sedna's. Our previous work had been limited to examining a model population of bodies residing specifically on Sedna's orbit. Any realistic Sedna population likely occupies a much larger region of orbital space, possibly including objects with sufficiently high perihelia that they would never or rarely become bright enough to see. With secure orbital classifications for survey objects, we can now test more sophisticated orbital distributions. 

\subsection{Constraints on a Cluster Birth}

No new Sedna-like bodies with perihelia beyond 45 AU were found in the survey despite a sensitivity out to distances of $\sim$1000 AU. Although, we cannot differentiate between the Sedna origin scenarios with a single detection, we can place constraints on the cluster birth model where the location and distribution of Sedna-like orbits is indicative of the SunÕs birth cluster size. Most stars are born in dense gas-rich embedded clusters \citep{1991ApJ...374..533L,2000AJ....120.3139C,2003AJ....126.1916P,2003ARA&A..41...57L,2007prpl.conf..361A}, and it is likely that the Sun spent several million years in such an environment. The presence of short-lived radioactive nuclides in primitive meteorites, may provide circumstantial evidence that the Sun was in relatively close proximity to a supernovae early on in the solar system's formation, (\citeauthor{2007CRGeo.339..872C} \citeyear{2007CRGeo.339..872C};\citeauthor{2009M&PSA..72.5303B} \citeyear{2009M&PSA..72.5303B}  and references therein) and therefore in a much denser environment than the present-day solar neighborhood. In the dense stellar nursery, encounters between nearby solar neighbors and the Sun would occur at a much higher frequency than in the present solar environment \citep{2001Icar..150..151A,1998ApJ...508L.171L,2009ApJS..185..486P,2010arXiv1001.5444A}. Close fly-bys of passing stars would perturb objects in the Sun's planetesimal disk onto highly eccentric Sedna-like orbits \citep{2004AJ....128.2564M,2006Icar..184...59B,2007Icar..191..413B,2008Icar..197..221K}. 

\begin{figure}
\epsscale{1}
\plotone{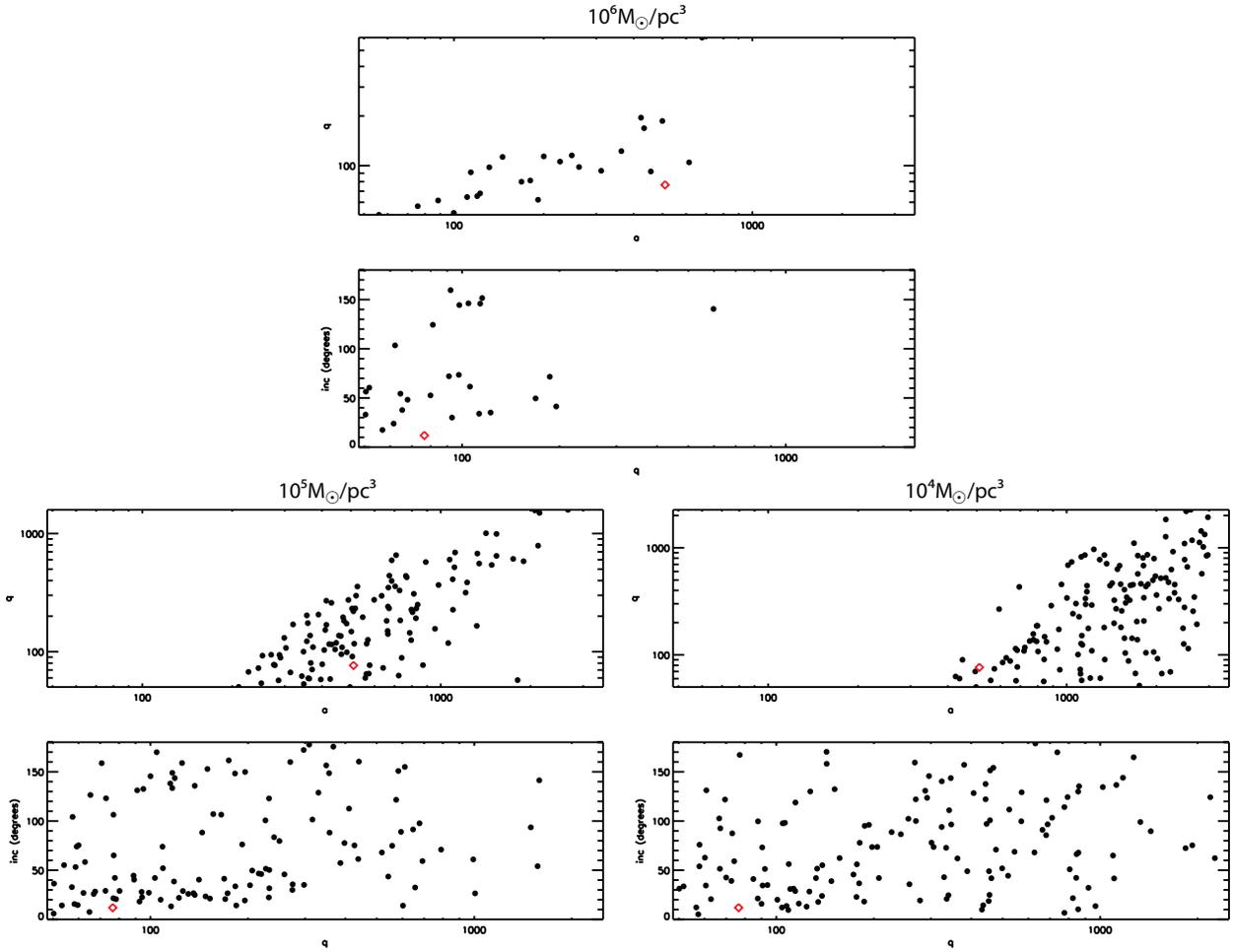}
\caption{Plot of perihelion (q) vs. semimajor axis (a) and plot of inclination (i) vs. semimajor axis (a) for Sedna-like bodies produced at the end of  the \cite{2006Icar..184...59B} embedded cluster simulations used in this work. We limit the population to orbits $q$ $>$ 50 and $a$ $<$ 3,000 AU. The diamond (red in the online version) denotes Sedna's orbit.}
\label{fig:brasser}
\end{figure}

\cite{2006Icar..184...59B} successfully produce objects on orbits similar to Sedna's in simulations of embedded cluster environments. The gravitational effects of both stars and gas in the cluster are included in their integrations. If the mean density of the material the Sun encounters while residing in the embedded cluster was  $\sim$$10^3$ M$_\odot$/pc$^3$ (central cluster densities of $10^4$ M$_\odot$/pc$^3$ ) or denser, Sedna's orbit is recreated and a distribution of Sedna-like bodies with semimajor axes less than 10,000 AU is formed. \cite{2006Icar..184...59B} find that the central density of the stellar cluster (directly correlated to the amount of material the Sun encounters in the cluster) determines the orbital distribution of Sedna-like bodies generated. The denser the cluster environment, the smaller semimajor axis at which the Sedna population begins. For this paper, we focus specifically on the \cite{2006Icar..184...59B} results for the $10^4$,$10^5,$ and $10^6$ M$_\odot$/pc$^3$ embedded cluster integrations ($10^3$ M$_\odot$/pc$^3$ did not produce Sedna). We refer the reader to their paper for details of the orbital integrations and the review of embedded clusters by  \cite*{2003ARA&A..41...57L}. Figure \ref{fig:brasser} shows the orbital distributions from the embedded cluster numerical simulations used in this work. 

 Our survey observations probe the Sedna population today after 4.5 Gyrs of evolution. The distribution of orbits presented by \cite{2006Icar..184...59B} is what remains after 3 Myr when the integrations end and the Sun is expected to have left the birth cluster. Once the Sun exits the cluster, the Galaxy becomes the dominant gravitational potential. The gravitational perturbations from galactic tides over the age of the solar system have not been accounted for in the \cite{2006Icar..184...59B} integrations. Sedna's orbit is protected from the effects of passing stars and galactic tides in the current solar environment, but objects with higher semimajor axes than Sedna may be perturbed onto comet-like orbits  \citep{1987AJ.....94.1330D,1997Icar..129..106F}.  \cite{2009Sci...325.1234K} examined the production of long period comets in the Sedna region and find that the production efficiency drops significantly for bodies with a $<$ 3000 AU compared to those with larger semimajor axes. Therefore we expect that objects emplaced onto Sedna-like orbits with semimajor axes less than 3000 AU should remain to the present day, and we do not include any orbits from the cluster simulations with a $\ge$ 3000 AU in comparisons to our observations.

The Nice model \citep{2005Natur.435..459T,2005Natur.435..466G} predicts that the giant planets were in a more compact configuration than in the present-day solar system. The orbits of the giant planets went unstable approximately 1 Gyr after the formation of the solar system causing the migration of the giant planets and scattering of planetesimal disk. Jupiter migrates inward, and the remaining giant planets move outward with Neptune migrating outward to 30 AU. The oldest embedded clusters are $\sim$5 Myrs old \citep{1989ApJS...70..731L,2003ARA&A..41...57L}, Neptune migrates well after the Sun has left the birth cluster and the emplacement of the Sedna population. \cite{2008A&A...492..251B} confirms this scenario can create a Sedna population and generate an Oort cloud population within the current estimates of the mass of the Oort cloud. Neptune's orbit became eccentric during migration and was later circularized via scattering of planetoids in the Kuiper belt region. Current estimates have Neptune's eccentricity as high as $\sim$0.3 corresponding to an aphelion of $\sim$39 AU at the end of migration \citep{2008Icar..196..258L}. The sculpting of the Sedna population due to Neptune's migration outward has not been accounted for in the  \cite{2006Icar..184...59B} simulation results. The cluster models do create orbits with perihelia in the range of 30-50 AU, which may not exist in the current solar system due to Neptune ejecting these Sednas or scattering them onto KBO-like orbits during itÕs eccentric phase. We chose a conservative minimum perihelia threshold of 50 AU (which would require Neptune to have an eccentricity of $\sim$0.7 to reach 50 AU at aphelion) to compare the cluster distributions to our survey results.  

\subsection{ Survey Simulator}
\label{sec:simulator}
We developed a survey simulator to compare the expected number of detections from the theoretical cluster Sedna populations to our survey results. The simulator draws synthetic objects from a model orbital and absolute magnitude distribution and for every image computes the positions and brightnesses of these objects on the sky. For all three cluster environments, we model a population of 3,000,000 bodies on cluster-created orbits randomly drawing the semimajor axis, eccentricity, and inclination for each particle from those produced in the cluster numerical integrations. \cite{2006Icar..184...59B} obtain a value of $\sim$2 Gyr for the precession frequency of Sedna and other Sedna-like objects, therefore we assume that the orbits have been randomized due to planetary effects, and randomize over all other orbital angles. The positions of the artificial objects are computed for the survey period; those synthetic cluster objects that land on our images are identified neglecting the effects of masked regions of the CCDs. For each of the three cluster environments, approximately a third of the synthetic cluster-created orbits are located on our images.

A brightness distribution is then applied to the synthetic population. Due to the large uncertainties in the albedo distribution of such a distant population, we assign absolute magnitudes to our synthetic bodies instead of diameters. We assume a single power-law brightness distribution similar to the Kuiper belt where the number of objects brighter than a given absolute magnitude, H$_{max}$, is described by:
\begin{equation}
 {\rm N} (H \le H_{max})=N_{H\leq1.6}10^{ \alpha (H_{max}-1.6)}
 \end{equation}
   The brightness distribution is scaled to $N_{H\leq1.6}$, the number of bodies with an absolute magnitude brighter than or equal to Sedna (H=1.6). A typical value of $\alpha$ measured for the asteroid belt is 0.3 \citep{1998Icar..131..245J}. The best-fit single power law for the hot (inclinations $> 5^\circ$) and cold (inclinations $< 5^\circ$)  Kuiper belt populations are $\alpha$=0.35  and $\alpha$=0.82 respectively, measured by Fraser et al (2010), but it is unclear if the Sedna population should have an $\alpha$ value similar to the Kuiper belt. The Sedna population may have very different surface characteristics than typical KBOs. \cite{2005A&A...439L...1B} find methane and a tentative detection of nitrogen on Sedna's surface. \cite{2007ApJ...659L..61S}  model of volatile loss on KBO surfaces predicts that moderate-sized Sedna-like bodies on high perihelia orbits should retain methane and nitrogen ices on their surfaces. Most KBOs on the other hand, are either too small or too hot to hold on to their primordial abundance of volatiles. Distant Sednas never sublimate a significant amount of ices to renew their surfaces in a frost/thaw cycle. Instead the surfaces of the Sedna population would be subject to constant photoprocessing of methane by solar irradiation steadily darkening their surfaces. Sedna is one of the reddest KBOs with a V-R=0.78 with thermal measurements constraining Sedna's V albedo to be between 0.16 and 0.30 \citep{2008ssbn.book..335B,2008ssbn.book..161S}. We choose to explore the extremes of the brightness distribution and model the likely range of power-law distributions for $\alpha$ ranging from 0.2-0.8 including the best-fit value for the hot and cold KBOs measured by Fraser et al (2010). 

For a given value of $\alpha$ and  $N_{H\leq1.6}$ absolute magnitudes are randomly assigned to our simulated Sednas. A single instance of the brightness distribution can be thought of as a separate survey. Those synthetic objects that lie within our sky coverage with an apparent magnitude above both nights' limiting magnitudes (as determined in Section \ref{sec:limitingmag}) are deemed valid survey detections. We assume a 100$\%$ efficiency out to the limiting magnitude where then the efficiency immediately drops to zero. We require that the object must be located on all 4 field images to be considered ``discovered'' in the simulated survey, and  we do not require the object have Sedna's perihelia of 76 AU.  Bodies with H $\leq$ 4.3 residing at 50 AU would be visible within our survey, and an object of Sedna's size and albedo would have been detected up to a distance of $\sim$93 AU. Objects have multiple detection opportunities due to repeat sky coverage over subsequent years and overlapping fields. We do not count duplicate detections in our tallies. 

\subsection{Could Sedna have been formed in a cluster environment?}

We did not find any distant objects with perihelia greater than 45 AU with the exception of Sedna. To determine whether the orbital distributions produced in the various cluster environments are consistent with our redetection of Sedna, we must compare the orbital distributions of single detections produced by the survey simulator to Sedna. We employ our 3,000,000 synthetic Sedna population for each cluster environment to generate single detections.  For each given value of $\alpha$, absolute magnitudes are randomly assigned to our simulated Sednas for the range of possible values of  $N_{H\leq1.6}$ to create 10,000 single detections. 

Each simulated detection is characterized by a semimajor axis, eccentricity, and inclination (\emph{ a,e,i}).  We test \emph{ a,e,i} because these three parameters are directly effected by the impulses from the stellar encounters and gravitational effects from the embedded gas and stars, and these are the most independent set of orbital parameters. We choose to exclude the H distribution in our analysis because of the uncertainty of our limiting magnitudes. To determine whether Sedna and the cluster produced single detections could be drawn from the same parent population for varying slopes and scaling of the brightness distribution, we employ a variant of  a 3-dimensional ( 3-D)Kolmogorov-Smirnov (KS) test adapted from  \cite{1983MNRAS.202..615P}  and \cite{1992nrca.book.....P} which simultaneously compares the \emph{ a,e,i} orbital distributions to Sedna. The fraction of data points in each of the 8 quadrants in $a,e,i$  space, where the origin is defined by Sedna's orbital parameters ($a$=519 AU , $e$=0.853, $i$=11.9 deg), is computed. In order to determine if Sedna's orbit is extreme compared to the cluster produced detections, the D statistic in this case is defined as the difference of the maximum and minimum fraction calculated. The significance of the computed D statistic is found by performing our 3-D KS test again, selecting each of the 10,000 simulated single detections as the new origin, counting the fraction where the computed D statistic was higher than the D statistic for Sedna's orbit.  We reject the cluster-produced population if the 3-D KS test does reject at  a 95$\%$ or greater significance  the null hypothesis, that the simulated survey single detections and our sole detection of Sedna are drawn from the same distribution.

We performed the 3-D KS test for all ranges of $N_{H\leq1.6}$ that produced single detections and possible values for $\alpha$ (0.2-0.82) for all three cluster environments. The orbital distribution of single detections produced at smaller $N_{H\leq1.6}$, is different from those at large $N_{H\leq1.6}$, and the entire range of possible values $N_{H\leq1.6}$ must be tested. At small values of $N_{H\leq1.6}$ there are fewer  bright H objects available to fill detectable orbits, biasing the single detections to  slightly lower perihelia orbits than for  larger values of $N_{H\leq1.6}$ where there is an ample supply of bright bodies to fill detectable orbits. We find that the  3-D KS test confidence levels calculated for the $10^6$ and $10^5$ M$_\odot$/pc$^3$cluster distribution for varying values of $\alpha$ are independent of $N_{H\leq1.6}$.  For the  $10^4$ cluster and any value of $\alpha$, $N_{H\leq1.6}$=1 has the highest probability of rejection and then decreases to a flat value as   $N_{H\leq1.6}$ increases. For the  $10^4$ cluster,  $N_{H\leq1.6}$=1 represents an upper limit on the rejection confidence level of the orbital distribution. Therefore we report  the confidence level calculated for each cluster distribution and brightness distribution for values of $N_{H\leq1.6}$=1 in Table \ref{tab:cluster}.  For the two densest cluster environments $10^6$ and $10^5$ M$_\odot$/pc$^3$ producing Sedna as the sole detection is an extremely low probability event. The bulk of the $10^6$ and $10^5$ M$_\odot$/pc$^3$ cluster-created single detections had orbits with  semimajor axes less than Sedna's. The simulations produce many more objects with lower perihelia than Sedna that should have been found but were not detected in our survey. We can rule out $10^6$ and the $10^5$ M$_\odot$/pc$^3$ cluster population at confidence levels greater than 95$\%$ for all ranges of $\alpha$ and possible values of $N_{H\leq1.6}$. Therefore we reject the $10^6$ and $10^5$ M$_\odot$/pc$^3$ clusters as the source of the Sedna population. We cannot reject the $10^4$ M$_\odot$/pc$^3$ cluster environments to a confidence level greater than 60$\%$ for all combinations of $\alpha$ and $N_{H\leq1.6}$ tested; we are unable to rule this population out with statistical significance. The $10^4$ M$_\odot$/pc$^3$ orbital distribution is consistent with our redetection of Sedna. These results assumed that  every object that lands on a CCD brighter than the image limiting magnitude would be detected. Our detection efficiency is not 100$\%$, but including a flat detection efficiency curve that drops to zero at the image limiting magnitude does not change the results presented. Including a detection efficiency produces the same types of orbits for single detections, just the absolute number of single detections decreases. Since we are only looking at single detections, the 3D KS test results are the same for any efficiency value. 

If Sedna's orbit is the result of multiple stellar encounters when the nascent Sun resided in an embedded cluster, our work rules out central densities for the cluster greater than or equal to $10^5$ M$_\odot$/pc$^3$ for the environment of the early solar system, and  \cite{2006Icar..184...59B} requires central densities higher than $10^3$ M$_\odot$/pc$^3$ to reproduce Sedna's orbit. In terms of the mean density of the material the Sun would have interacted with in the cluster environment,  the Sun would have had to have encountered a mean density greater than $10^3$ and less than $\sim10^4$ M$_\odot$/pc$^3$ to be consistent with our survey observations. \cite{2005ApJ...632..397G} map the volume density of three young embedded cluster regions (GGD 12-15, IRAS 20050+2720, NGC 7129). The peak densities of these regions were on the order of $\sim10^5$ M$_\odot$/pc$^3$.  For GGD 12-15 and IRAS 20050+2720, 72$\%$ and 91$\%$ of the member stars reside in locations with densities upwards of $10^4$ M$_\odot$/pc$^3$, and are unlikely to produce the observed Sedna population.  For NGC 7129, less than 24$\%$ of the stars in the core of the cluster experience densities greater than $10^4$ M$_\odot$/pc$^3$. \cite{1995AJ....109.1682L} estimate the central stellar density of the 0.1 pc central regions of IC 348, NGC 2024, and Trapezium clusters to range from $\sim 10^3-10^4$ M$_\odot$/pc$^3$3 at the minimum central density required to form Sedna's orbit. These environments and NGC 7129 could produce the observed Sedna population. 

\subsection{Population Estimate}
\label{sec:sednasize}
Now that we have found the $10^4$ M$_\odot$/pc$^3$ cluster population is the only cluster environment capable of emplacing Sedna on itÕs orbit, we can place constraints on the size of the produced population. To estimate the size of the Sedna population, we use the value of $\alpha$ measured by Fraser et al (2010) for the hot and cold populations of the Kuiper belt  ($\alpha$=0.35 and 0.82 respectively) as limits for our brightness distribution. For each given value of $\alpha$, absolute magnitudes are randomly assigned to our survey simulator created 3,000,000 Sednas  50,000 times, for every value of  $N_{H\leq1.6}$. A single instance of the brightness distribution can be thought of as a separate survey. For each  $N_{H\leq1.6}$ tested, the number of synthetic ``surveys'' in which, like the real survey, one object on a Sedna-like body is detected are tallied.    Valid detections are only those in which the object is located on the same CCD and in all 4 field observations. We do not require that the object have Sedna's absolute magnitude (H=1.6), only that the apparent magnitude of the object is above the SExtractor calculated limiting magnitudes  of  all 4  frames the object is ``discovered''on. 

\begin{figure}
\epsscale{.75}
\plotone{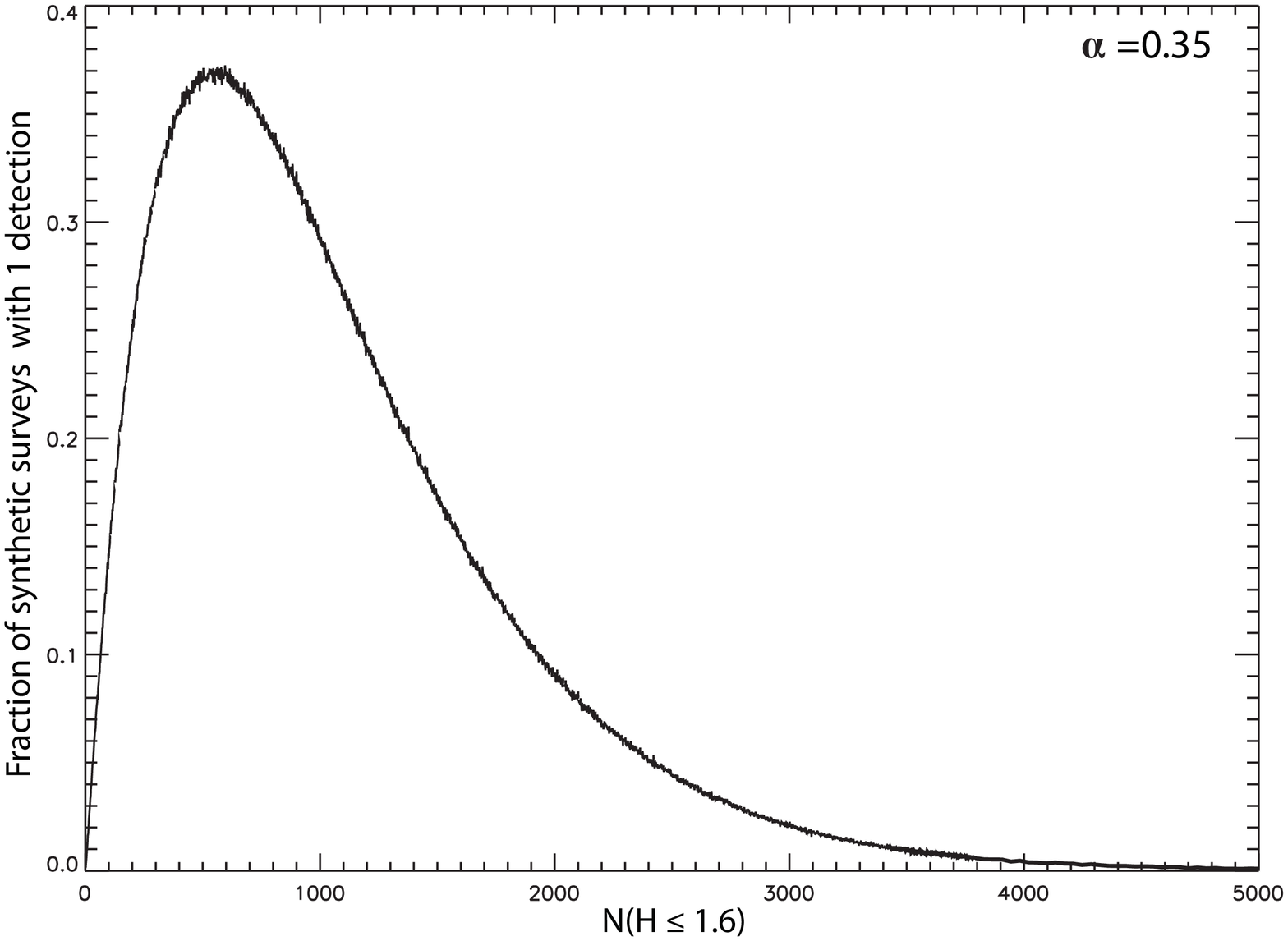}
\plotone{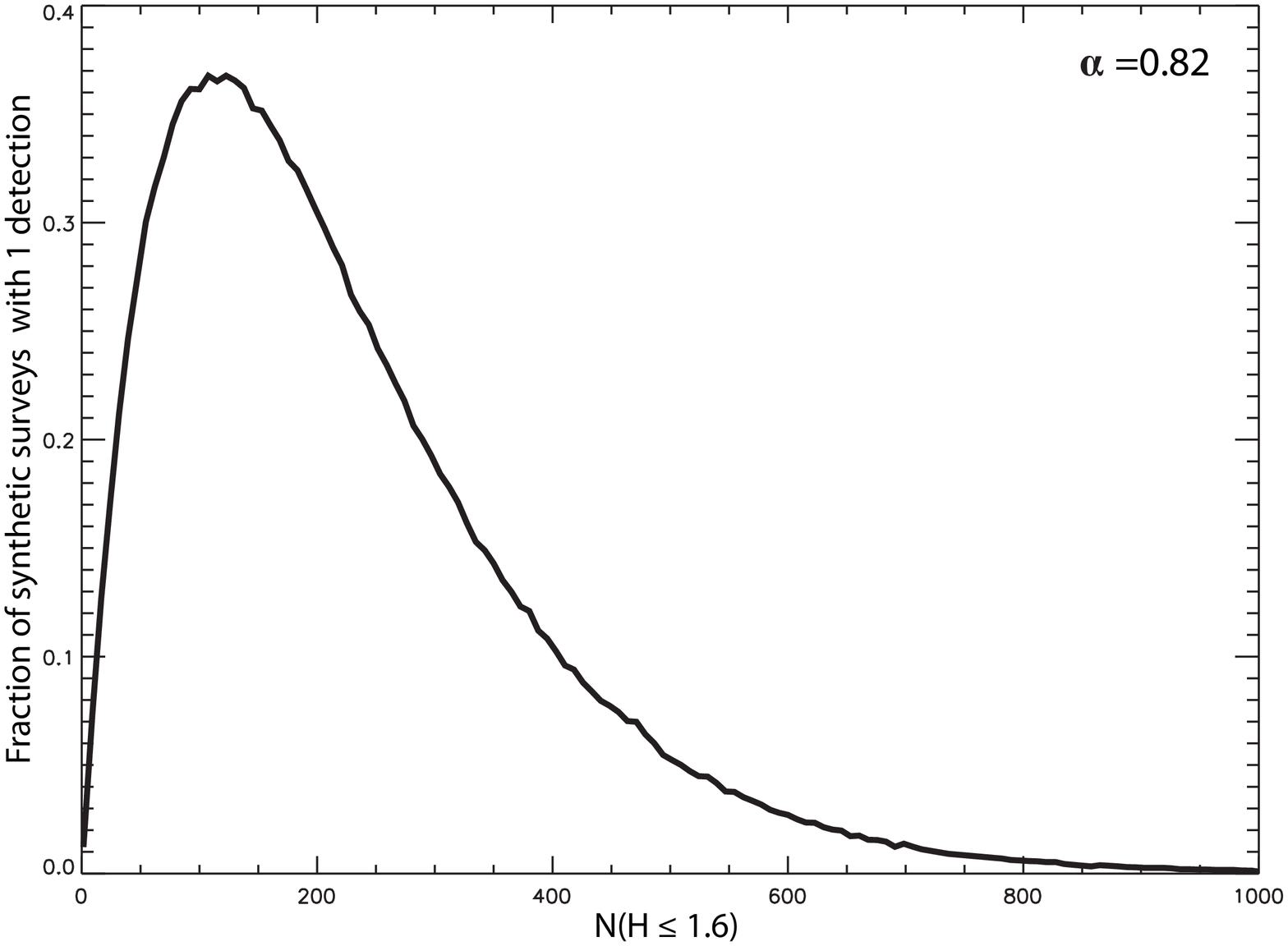}
\caption{Results of the 10$^4$ M$_\odot$/pc$^3$  cluster analysis for the $\alpha$=0.35 (hot)  and $\alpha$=0.82 (cold) Kuiper belt population size distributions- Fraction of synthetic surveys with one detectable Sedna-like body as a function of the number of bodies bigger and brighter than Sedna assuming our nominal 66$\%$ detection efficiency.}
\label{fig:cluster}
\end{figure}

The best-fit values for the number of objects brighter than or equal to Sedna with 95$\%$ errors are 393 $^{+1286}_{ -264}$ and 74$^{+ 279}_{-47}$ for the hot and cold brightness distributions respectively. The lower and upper 95$\%$ confidence levels limits reported are one-sided statistics found by computing the interval over which the integrated probability distribution 0.95 respectively of the total area. The survey simulator assumes all simulated Sednas that land on our images and are above the image limiting magnitude would be detected in the survey. The effect of a less than 100$\%$ survey detection efficiency is non-negligible. The reported size estimates represent  a lower-bound on the size of the Sedna population. Assuming a uniform detection efficiency which drops to zero at the image limiting magnitude, the  best-fit value and 95$\%$ limits for $N_{H\leq1.6}$  is scaled by the inverse of the survey efficiency.  For our nominal detection efficiency of 0.66, the best-fit values for the number of objects brighter than or equal to Sedna are 595$^{+ 1949}_{-400}$ and 112$^{+ 423}_{-71}$  respectively for the hot and cold brightness distributions. Figure \ref{fig:cluster} plots the fraction of simulated surveys that produced a single Sedna detection as a function of  $N_{H\leq1.6}$ the $10^4$ M$_\odot$/pc$^3$ cluster environment for our nominal survey detection efficiency. 

For the  $10^4$ M$_\odot$/pc$^3$ cluster environment, the range is quite large but there could be on the order of hundreds to thousands of planetoids brighter than Sedna present beyond the Kuiper belt. For comparison, the total number of Sedna-sized or larger bodies in the Kuiper belt is $\sim$5-8 \citep{2008ssbn.book..335B}; there may be an order of magnitude or two more mass residing in the Sedna region than exists in the present Kuiper belt. The expected number of objects with H$\le$1.6 varies significantly with the slope of the brightness distribution. Choosing a steeper power law for the brightness distribution decreases the likelihood of detecting only one Sedna because of the larger number of bright objects populating detectable orbits and decreases the best-fit number of objects brighter than Sedna. Selecting a smaller value of $\alpha$, a shallower brightness distribution, increases the likelihood of detecting only one object on Sedna's orbit by decreasing the number of synthetic surveys with multiple detections.  

We excluded orbits with semimajor axes greater than 3000 AU from our analysis. If we had included higher semimajor axis orbits, the best-fit number of Sedna-like bodies brighter than or equal to Sedna would increase due to the larger population of orbits being included, but our overall conclusions would not change. Those orbits that are contributing most to being Sedna detections are those with semimajor axes much smaller than 3000 AU. The $10^4$ M$_\odot$/pc$^3$ cluster results are the most sensitive to this cut. We examined the single detections produced in the best-fit simulations for the range of $\alpha$ parameters. Orbits with semimajor axes less than 1000 AU contribute $\sim$80-90$\%$ of the Sedna single Sedna detections. The number is even greater from the $10^6$ and $10^5$ M$_\odot$/pc$^3$clusters. We make a conservative perihelia cut at 50 AU to compare the cluster produced Sedna-like orbits to our observed Sedna population. Our simulations were rerun, including orbits with perihelia greater than 45 AU and our conclusions remain the same. The $10^6$ and $10^5$ M$_\odot$/pc$^3$ cluster will be ruled out with a higher confidence level because of the increase in low perihelia orbits that should have been detected in our survey. For this analysis we used the SEXtractor computed limiting magnitudes to determine whether our simulated Sedna population was observable on our images. 

\subsection{Comparison to Occultation Surveys}

\cite{2009AJ....138.1893W} place upper limits on the number of small bodies in the Sedna region due to the lack of occultations from distant solar system bodies in the TAOS survey. We can estimate whether our upper limits for the Sedna population are consistent with \cite{2009AJ....138.1893W}'s reported upper limits on the number density of objects  larger than 1 km. We find the fraction of our $10^4$ M$_\odot$/pc$^3$ cluster survey simulator created 3,000,000 Sednas that are within 3$^\circ$ of the ecliptic (TAOS's ecliptic latitude range) and at 100 and 1000 AU. Assuming no break in the size distribution, we extrapolate the number of bodies larger than 1 km. The albedo distribution is uncertain, Sedna's V albedo is measured to be between 0.16 and 0.30 \citep{2008ssbn.book..335B,2008ssbn.book..161S}, but to give an extreme upper limit we chose an albedo for 0.04 and assume no break in the size-distribution in order to estimate the fraction of bodies that would be observable by TAOS. TAOS is sensitive to bodies brighter than H=19.1. 

Within 3$^\circ$ of the ecliptic, 0.05$\%$ of the 10$^4$ M$_\odot$/pc$^3$ cluster produced Seda population are located between 50-150 AU  and 0.07$\%$ reside at 900-1100 AU. For the flat size distribution value ($\alpha$=.35), we expect there to be no more than 780 Sednas/deg$^2$ on the ecliptic at 100 AU and 10$^4$  Sednas/deg$^2$ at 1000 AU assuming a 66$\%$ detection efficiency. Our 95$\%$ confidence level  estimates for a flat size distribution are well below TAOS's ecliptic number density of 1 km or larger bodies at 100 ($\sim$10$^7$  Sednas/deg$^2$) and 1000 AU ($\sim$10$^9$  Sednas/deg$^2$) even without a break in the size distribution. The TAOS observations do not rule out a large Sedna population with thousands of Sedna-sized or larger bodies residing far from the Sun for a flat brightness distribution. For the steep (cold population) size distribution, $\alpha=.82$, and a 66$\%$ magnitude detection efficiency, at 100 AU we expect no more than 2.8x$10^{10}$ objects/deg$^2$, approximately two orders of magnitude larger than TAOS's 3-$\sigma$ upper limit. We find that even our 95$\%$ lower limit at 100 AU is an order of magnitude larger than the TAOS limit. Our expected number density at 1000 AU at our 95$\%$ upper confidence level,  on the other hand, is  3.70x$10^{11}$  objects/deg$^2$ below TAOS's limit of $\sim$10$^{12}$  objects/deg$^2$. 

The occultation results do not necessarily rule out a steep size distribution for the Sedna population. In the Kuiper belt at small sizes ($\sim$50-150 km) the distribution is observed to break to a shallower slope \citep{2004AJ....128.1364B,2009ApJ...696...91F,2009AJ....137...72F}.  \cite{2006Icar..184...59B}'s model did not include gas in the solar nebula and therefore did not include the effects of gas dynamics in their simulations.  Sedna is $\sim$1500 km in size and would not be effected by gas drag, but smaller sized objects would be. \cite{2007Icar..191..413B} investigated the effect of gas drag on the size distribution of objects deposited into the Sedna region. They find a size sorting effect in the cluster-produced Sedna population. Bodies smaller than $\sim$20-60 km would be circularized onto orbits beyond Jupiter and Saturn and not available to be scattered into the Sedna region. Far fewer small-sized objects would be deposited into the Sedna region. Our survey is sensitive to objects much larger than those that would be effected by gas drag or the break in the brightness distribution. The combination of a broken power-law size distribution and a size-sorting effect could reconcile the observations, causing very few small objects that TAOS would have been able to detect to be present in the Sedna region.

\subsection{Open Cluster Environments}

The majority of stars are birthed in embedded clusters, but 4-7$\%$ of stars form in smaller loose conglomerations with little or no gas known as open clusters \citep{2004ASPC..323..161L}.  Open clusters, like the Pleiades, have ages of a few tens to hundreds of Myrs \citep{2004ASPC..323..161L}. Although embedded clusters are more prevalent, it is postulated that $\sim$5$\%$ of the embedded clusters may dissipate into loosely bound open clusters \citep{2003ARA&A..41...57L}.  \cite{2008Icar..197..221K} are able to produce objects on Sedna-like orbits in various open cluster environments. Interactions between the planetesimals disks of the cluster members are not included in their simulations. Their numerical integrations produce similar wedge-like orbital distributions to the  \cite{2006Icar..184...59B} embedded clusters models, but  \cite{2008Icar..197..221K} find no relationship between the size of the birth cluster and the orbital distribution of Sednas. The open cluster integrations are nondeterministic with Sedna's orbit being produced in only 5 of their 16 cluster simulations of varying cluster size. For those integrations that do produce Sedna and other Sedna-like orbits, distributions similar to the $10^4$ and $10^6$ M$_\odot$/pc$^3$  \cite{2006Icar..184...59B} results are generated. This is not unsurprising since the dominant dynamics sculpting the Sedna region, stellar encounters, is the same in both environments. Our analysis above of the embedded cluster distributions also applies to \cite{2008Icar..197..221K} open cluster orbital distributions. Those distributions where Sedna is at the end of a distribution Sedna-like orbits with many lower semimajor axes and lower perihelia orbits similar to the $10^5$ and $10^6$ M$_\odot$/pc$^3$3 embedded clusters, are inconsistent with our observations. 

\subsection{Implications for the Kuiper Belt}

Using the discovery of 2008 KV42, with an orbit essentially perpendicular to the ecliptic, \cite{2009ApJ...697L..91G} posit a metastable parent population with inclinations greater than $\sim$50 AU with $a$ in the hundreds of AU and $q= 35$Ð$45$ AU. Such a population is produced in the $10^5$ and $10^6$ M$_\odot$/pc$^3$ cluster environments but not present in the $10^4$ M$_\odot$/pc$^3$ embedded cluster \citep{2006Icar..184...59B}. \cite{2009ApJ...697L..91G} suggest 2008 KV42 may have been a high inclination counterpart to Sedna placed on a lower perihelia and semimajor axis that later diffused to itÕs current orbit. Although for our analysis we removed such objects with perihelia less than 50 AU from our distribution, adding those objects would only rule out the $10^5$ and $10^6$ M$_\odot$/pc$^3$ cluster environments to even higher confidence because many more low $a$ and low $q$ objects single detections would be produced than detections with similar orbits to Sedna. With this region devoid of particles in the $10^4$ M$_\odot$/pc$^3$ cluster integrations, this suggests that 2008 KV42 and Sedna are likely formed from two independent source populations.

\section{Latitude Distribution}
\label{sec:integrations}

\begin{figure}
\epsscale{1}
\plotone{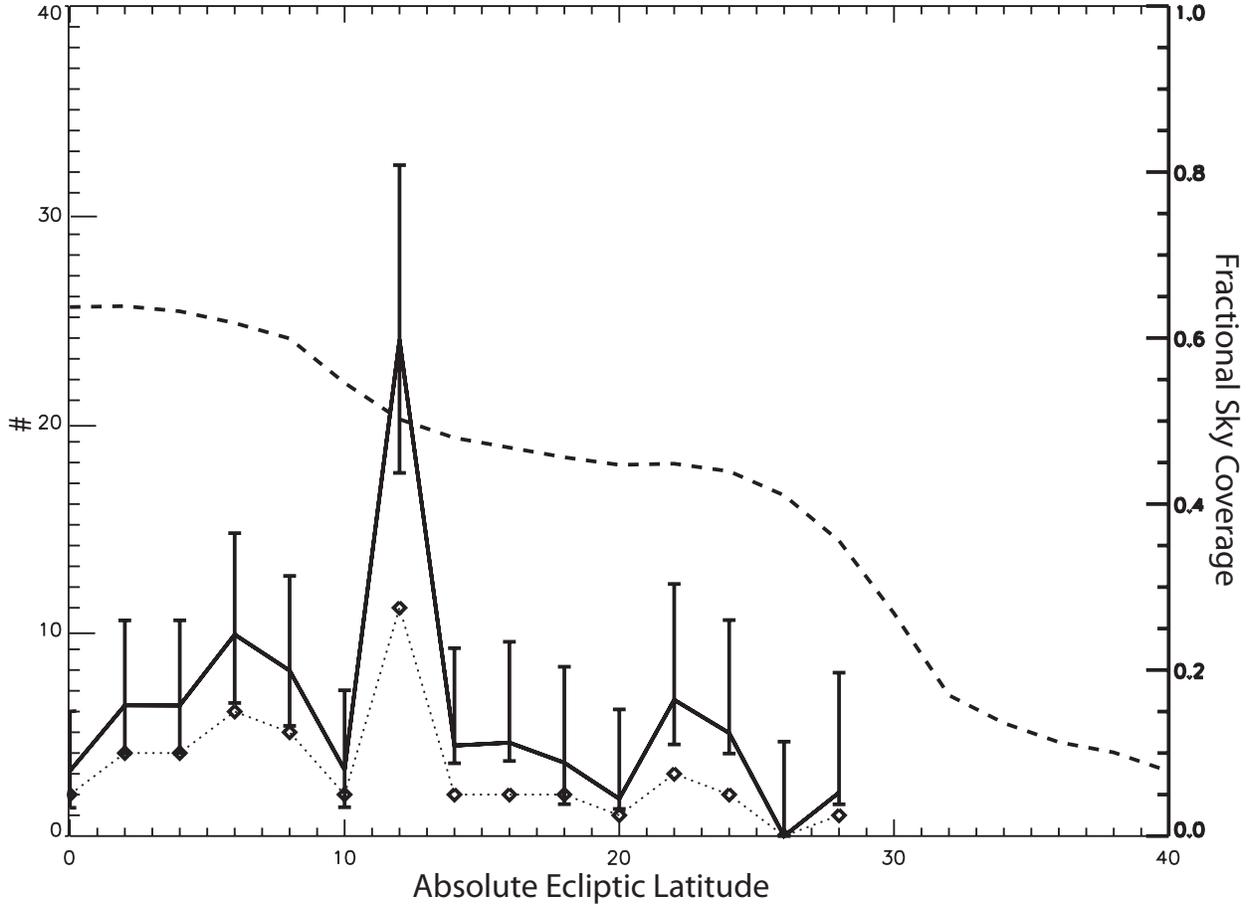}
\caption{The folded latitudinal distribution of objects with semimajor axis $>$ 30 AU found in this work. The lower dashed line with diamonds shows the number of actual KBO detections in two-degree bins. The dashed line shows the fractional ecliptic latitude completeness. The solid line shows the expected number of KBOs brighter than 21.3 corrected for sky coverage with one-$\sigma$ Poisson error bars computed for the unfolded distribution added in quadrature.}
\label{fig:lat}
\end{figure}

Figure \ref{fig:lat} plots the folded latitude distribution of all objects with a$>$30 debiased for latitudinal coverage. We assume Poisson detection statistics (as computed by \cite{1991ApJ...374..344K}), with error bars representing the Poissonian 68$\%$ confidence limit on the detected number of objects in each latitude bin corrected for sky coverage. A noticeable spike occurs at  $\sim$12$^\circ$ from the ecliptic. \cite{2008ssbn.book..335B} also finds these prominent peaks in the latitudinal distribution $\sim$$\pm$ 11$^\circ$ ecliptic latitude.  Brown finds that this peaked distribution cannot be generated by a simple inclination distribution of objects in random orbits. \cite{2008ssbn.book..335B} suggests that resonant orbits are likely able to explain these high latitude concentrations. Resonators trapped in the Kozai resonance (such as Pluto) have their perihelia near their maximum excursion off the ecliptic \citep{1997Icar..127....1M} and the highest detection probability out of the ecliptic plane. The plutinos come to perihelia away from Neptune\citep{1996AJ....111..504M,1995AJ....110..420M} and are preferentially biased towards detection at certain longitudes. Without dynamical classification \cite{2008ssbn.book..335B} could not verify the plutinos as the source of these peaked latitude distributions. 

With secure orbits for our detections we can address this issue. In order to classify which of the survey KBOs reside in mean motion resonances with Neptune, each KBO had 13 clones integrated for 10 MYrs. One clone represents the best-fit orbit, and the rest are taken from a self-consistent spread of orbits covering the 3-$\sigma$ uncertainty of the KBOÕs best-fit orbital solution computed from the covariance matrix of orbital elements obtained from AstDys\footnote{\texttt{http://hamilton.dm.unipi.it/astdys/}} on 2009 December 1. These objects were integrated using the n-body code SyMBA \citep{1994Icar..108...18L} using the integrator \verb+swift_rmvs3+ based on the mapping by \citet{1991AJ....102.1528W}. The KBO clones were treated as massless particles. The four giant planets were included and their initial conditions were taken from JPL HORIZONS. The mass, position, and velocity of the terrestrial planets were combined with the Sun. The integration proceeded backwards in time with 40-day time steps from epoch JD 2455200. After 10 MYrs, the clones were examined for one or more librating resonant angles and as well as librating arguments of perihelion in order to identify Kozai resonators. We identified objects (listed in Table \ref{tab:kbos}) as resonant if all the clones lie in the resonance at the end of the integrations. 

\begin{figure}
\epsscale{.75}
\plotone{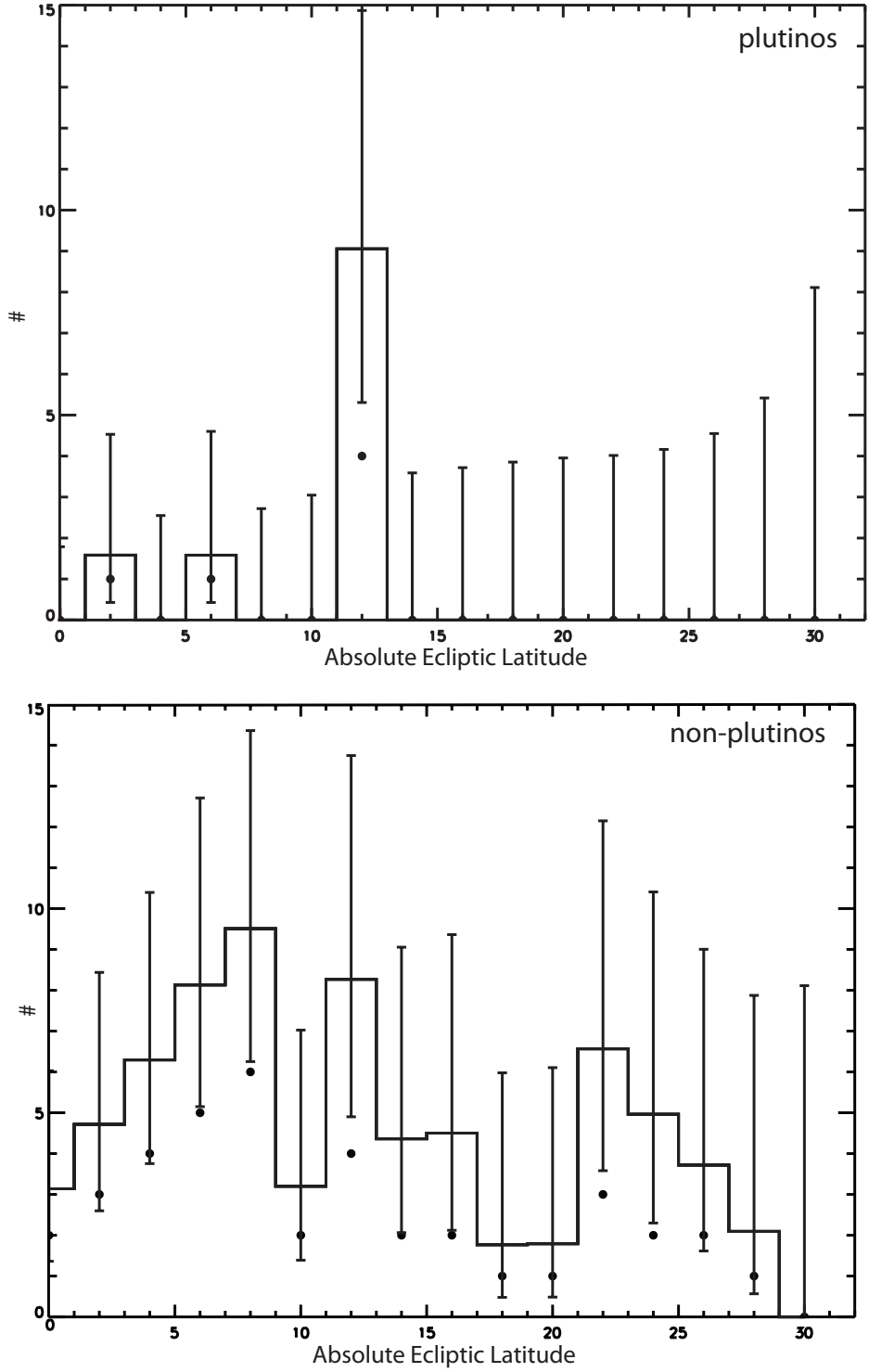}
\caption{The folded latitudinal distribution two-degree bins of survey of multi-opposition plutinos (top) and non-plutinos excluding confirmed Haumea family members (bottom) brighter than 21.3 corrected for sky coverage with one-$\sigma$ Poisson error bars computed for the unfolded distribution added in quadrature. Filled circles shows the number of actual KBO detections in two-degree bins.}
\label{fig:plutinos}
\end{figure}

The latitudinal distribution of detected plutinos found in the survey is plotted in Figure \ref{fig:plutinos}. Six plutinos were detected in our survey, only two reside at ecliptic latitudes less than 10$^\circ$. The remaining four plutinos compose the majority of the 12$^\circ$ latitude spike. Of these four plutinos (Huya, 2007 RT15, 2002 VE95 and 2008 SO2006), two objects are Kozai resonators, Huya and 2007 RT15; the other three have perihelia off the ecliptic having possibly experienced temporary Kozai interactions. The remaining non-plutino distribution still exhibited a peak in the distribution at 12$^\circ$ including two members of the Haumea collisonal family. At least 7$\%$ of our detections are fragments of the Huamea collisional family \citep{2007Natur.446..294B,2007AJ....134.2160R,2008ApJ...684L.107S,2010A&A...511A..72S}. The identifier of the Haumea family is the characteristic deep near-infrared pure water ice absorption features on their surfaces \citep{2007Natur.446..294B,2008ApJ...684L.107S}. The water ice-rich bodies are thought to all have anomalously high albedos, like family member  2002 TX300  \citep{2010Natur.465..897E}, extremely biasing our survey toward detection of Haumea family members. Any clustering in the Haumea family members will severely bias our latitude distribution. Removing the spectroscopicly confirmed family members from our survey, the non-plutino distribution is not peaked as shown in Figure \ref{fig:plutinos}. 

\cite{2008ssbn.book..335B} and this work are the only two wide-field surveys to probe significantly beyond the ecliptic. In order to test whether the plutino population observed by ecliptic surveys is representative of the entire plutino population, we compare our observed plutino latitude distribution to the CFEPS plutino model. The CFEPS survey \citep{2009AJ....137.4917K,2010G} orbital and brightness distribution is based on the sample of plutinos detected in observations covering ecliptic latitudes less than 2$^\circ$. None of their detections are Kozai librators, thus only representing the non-Kozai plutino population. CFEPS is sensitive to an absolute magnitude range of  H$_{g^\prime}$$\sim$ 6-10.5, fainter than the sources we are able to detect in our survey. In order to compare their model to our observed latitude distribution, we must extend the distribution to larger objects where the  CFEPS survey does not measure directly and where the slope of their measured brightness  distribution may not be applicable to the larger bodies that we detect. The H distribution is measured in g$^\prime$ and we observe in a broadband R filter. \cite{2008Icar..195..827F}  find an average KBO value of  $<$ g$^\prime$-R $>$= 0.95, and we apply this as our constant offset to the g$^\prime$ magnitudes. We create a latitude distribution by shuffling the absolute distribution of the $10^5$ model plutinos with H$_g$$>$10.5 $\sim$$10^5$ times. We tally the latitudes of all plutinos for all runs with magnitudes brighter than R=22 that lie within our survey sky coverage in a folded latitude histogram binned in 2$^\circ$ latitude bins.  To estimate the expected number of plutinos in the 12$^\circ$ bin, we scale CFEPS model latitude distribution to the value of our folded latitude distribution at 2$^\circ$, the lowest latitude binned plutino detection in our survey. Assuming Poisson errors and using the quadrature of the fractional errors. 6.7 $_{-6.3}^ {+15.6}$ (68$\%$ confidence level) times as many plutinos reside in 11-13$^\circ$ from the ecliptic than are predicted by the non-Kozai plutino CFEPS model. Although the range is quite high, our latitude distribution suggests that the plutino population in particular the Kozai population has been underestimated and may be much larger than previous KBO surveys have reported. 

\section{Number of Bright Objects}
\begin{figure}
\epsscale{1}
\plotone{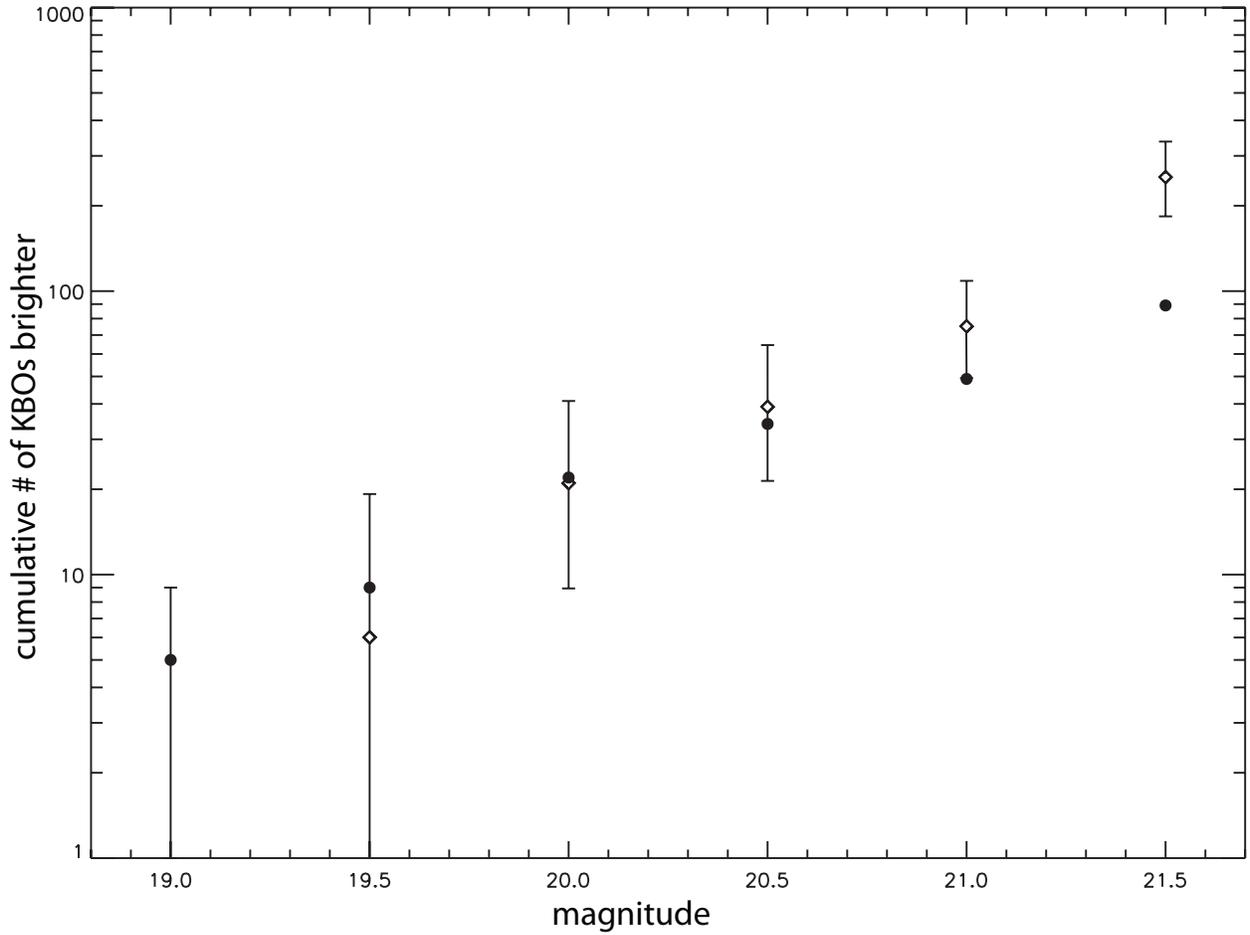}
\caption{Cumulative number of expected KBOs within $\pm$ 30$^\circ$ of the ecliptic assuming a flat latitude distribution (open diamonds) with 2-$\sigma$ Poisson error bars. Cumulative number of known multiopposition KBOs (a$\ge$ 30)within $\pm$30$^\circ$ ecliptic latitude (filled circles)}
\label{fig:numbright}
\end{figure}

Our survey probes the bright-end of the KBO size distribution. Assuming a uniform latitude distribution we can crudely estimate the number of large observable KBOs. Using our nominal survey efficiency function from Section \ref{sec:se}  and the effective area covered we compute the expected number of bright KBOs ($a$$>$30) as a function of magnitude. We neglect the effects of masked CCD regions and other geometric effects. Our sky coverage drops significantly above latitudes of $\pm$30$^\circ$ and therefore we only focus on detections and sky covered within $\pm$30$^\circ$ of the ecliptic. Figure \ref{fig:numbright} plots the cumulative number of expected bright KBOs as a function of magnitude compared to known multiopposition KBOs.    
 
 \section{Conclusions}

Surveying $\sim$12000 deg$^2$ within $\pm$30$^\circ$ of the ecliptic to $\sim$21.5 in R magnitude, we have searched for additional members of the Sedna population. Based on the 52 KBOs and Centaurs detected in our survey we conclude:
\begin{itemize}

\item We detected only one object on a Sedna-like orbit, Sedna, despite a sensitivity to motions of bodies out to $\sim$1000 AU. With one detection, we cannot differentiate between the various proposed formation mechanisms proposed to emplace Sedna on its orbit. 

\item For the embedded cluster Sedna formation model, we reject the $10^5$ and $10^6$ M$_\odot$/pc$^3$ cluster environment-produced populations as consistent with our redetection of Sedna.  We find the $10^4$ M$_\odot$/pc$^3$ cluster environment consistent with our observations, with a best-fit population of $N_{H\leq1.6}$= 595$^{+ 1949}_{-400}$  for the hot population and 112$^{+ 423}_{-71}$  for the cold population size distributions assuming our nominal detection efficiency of $66\%$.

\item The plutino population has a peaked distribution at $\sim$$\pm$12$^\circ$ ecliptic latitude, likely due to Kozai resonators and current estimates of the size of the plutino population from on-ecliptic surveys insensitive to these high latitude plutinos likely underestimate the size of the true population

\end{itemize}
 
{\it Acknowledgments:}
This research is supported by NASA Origins of Solar Systems Program grant NNG05GI02G. M. E. S. is supported by a NASA Earth and Space Science Fellowship. We are indebted to Ramon Brasser and Nathan Kaib for sharing the results of their cluster integrations and to JJ Kavellaars,Brett Gladman,and Samantha Lawler for providing us with the nominal CFEPS plutino model. We thank the staff at Palomar Observatory for their dedicated support of the robotic operation of the Samuel Oschin telescope and QUEST camera. The authors would also like to thank Greg Aldering for his help in scheduling the observations. We acknowledge Mansi Kasliwal, Henry Roe, John Subasavage, Emily Schaller, and Richard Walters for their assistance with recovery observations of our new discoveries. We recognize the work of Christian Clanton for support in developing the dynamical integration tools. We also thank Wes Fraser, Ramon Brasser, and Nathan Kaib for insightful conversations.  

{\it Facilities:} \facility{PO:1.2m}

\bibliographystyle{apj}

\begin{landscape}
\begin{longtable}{cllccccccc}
 \caption[short]{ Summary of Field Positions } \\

\hline
\hline
Pointing & \multicolumn{1}{c}{R.A. } & \multicolumn {1}{c}{Dec.} & \multicolumn{3}{c}{Night 1} & & \multicolumn{3}{c}{Night 2} \\ 
\cline{4-6}   \cline{8-10}
& \multicolumn{1}{c}{(J2000)}& \multicolumn{1}{c}{(J2000)}& MJD obs1 &  MJD obs 2 & mag. limit &  & MJD obs 1 & MJD obs 2 & mag limit  \\ \hline 

\endfirsthead

\multicolumn{3}{c}{{\tablename} \thetable{} -- Continued} \\[0.5ex]
  \hline \hline \\[-2ex]
Pointing & \multicolumn{1}{c}{R.A. } & \multicolumn {1}{c}{Dec.} & \multicolumn{3}{c}{Night 1} & & \multicolumn{3}{c}{Night 2} \\ 
\cline{4-6}   \cline{8-10}
& \multicolumn{1}{c}{(J2000)}& \multicolumn{1}{c}{(J2000)}& MJD obs1 &  MJD obs 2 & mag. limit &  & MJD obs 1 & MJD obs 2 & mag limit  \\ \hline 
\endhead

 \hline
  \multicolumn{3}{l}{{Continued on Next Page\ldots}} \\
\endfoot

  \\[-1.8ex] \hline \hline
  \caption[]{Full table can be found in the  online version. The center coordinates for all pointings searched for KBOs and used in the analysis presented in this paper. The table includes pointing number, the right ascension and declination of the mosaic center CCD (B15),  MJD dates of all four observations of the field and limiting magnitudes for each night the field was observed}  \\
\endlastfoot
           1 &  13 53 27.599 & -18 15 59.80 & 54228.327 & 54228.333 & 20.9 & & 54229.189 & 54229.237 & 21.4 \\
           2 &  13 55 33.959 & -18 15 59.80 & 54228.330 & 54228.336 & 21.1 & & 54229.192 & 54229.240 & 21.5 \\
           3 &  13 22 04.440 & -13 43 28.20 & 54228.313 & 54228.320 & 20.6 & & 54229.163 & 54229.209 & 20.9 \\
           4 &  13 38 45.960 & -13 43 28.60 & 54228.260 & 54228.266 & 21.4 & & 54229.169 & 54229.216 & 21.3 \\
           5 &  13 40 49.440 & -13 43 27.50 & 54228.263 & 54228.269 & 21.4 & & 54229.172 & 54229.219 & 21.4 \\
           6 &  13 55 27.841 & -13 43 27.80 & 54228.205 & 54228.211 & 21.2 & & 54229.176 & 54229.222 & 21.3 \\
           7 &  12 51 16.560 & -09 10 55.60 & 54228.273 & 54228.279 & 21.0 & & 54229.183 & 54229.230 & 21.4 \\
           8 &  12 53 18.240 & -09 10 55.90 & 54228.276 & 54228.283 & 21.1 & & 54229.186 & 54229.233 & 21.4 \\
           9 &  13 07 40.440 & -09 10 55.90 & 54228.219 & 54228.226 & 21.3 & & 54229.196 & 54229.243 & 21.5 
\label{tab:B15}
\end{longtable}
\end{landscape}

\begin{longtable}{|r c c c c c c c c c|}
\hline
 
designation & a  & e& i & R  & oppositions & night 1&  night 2 & H &  MMR \\ 
& (AU) & & (deg) & (AU) & & avg mag & avg mag &  &  \\ \hline
\endhead
 \hline
  \multicolumn{3}{l}{{Continued on Next Page\ldots}} \\
\endfoot

\hline 
  \caption[]{Orbital elements reported by the Minor Planet Center of Centaurs and KBOs detected in the Palomar survey: semimajor axis (a) , eccentricity (e) , inclination (i), barycentric distance (R), oppositions observed (in years excepted where days noted by d), nightly discovery magnitudes, absolute magnitude (H), and mean motion resonance (MMR) if applicable. All 2007 and 2008 classified objects were new discoveries found by this work} \label{grid_mlmmh} \\
\endlastfoot
(26181)  1996 GQ21& 93.01&0.588&13.4& 40.87& 11&21.0 & 20.6&  5.2  & 11:2
\\ 
(26308) 1998 SM165& 47.99&0.375&13.5& 36.99& 12&21.4 & 21.4&  5.8  & 2:1 Kozai
\\ 
(19521) 1998 WH24& 45.74&0.103&12.0& 41.84& 11&21.9 & 21.9&  4.8  &
\\ 
(40314) 1999 KR16& 48.83&0.306&24.8& 36.27&  6&20.6 & 20.7&  5.8  &
\\ 
(38628) 2000 EB173& 39.44&0.277&15.5& 28.93&  7&19.3 & 19.4&  4.7  & 3:2 Kozai
\\ 
(47932) 2000 GN171& 39.36&0.281&10.8& 28.34&  9&20.3 & 20.4&  6.0  & 3:2
\\ 
(20000) 2000 WR106& 42.85&0.056&17.2& 43.39& 13&19.9 & 19.7&  3.6  &
\\ 
(82075) 2000 YW134& 57.61&0.287&19.8& 43.81&  6&21.0 & 21.1&  4.9  & 8:3
\\ 
(83982) 2002 GO9& 19.45&0.277&12.8& 15.00&  6&21.1 & 21.1&  9.1  &
\\ 
(50000) 2002 LM60& 43.47&0.039& 8.0& 43.27& 16&19.3 & 19.4&  2.5  &
\\ 
(55636) 2002 TX300& 43.46&0.126&25.8& 41.29& 12&19.8 & 19.7&  3.3  &
\\ 
(55638) 2002 VE95& 39.37&0.290&16.3& 28.24& 10&20.1 & 20.0&  5.3  & 3:2
\\ 
(119979) 2002 WC19& 47.80&0.260& 9.2& 42.95&  7&21.4 & 21.4&  5.0  & 2:1
\\ 
(174567) 2003 MW12& 45.87&0.144&21.5& 47.95& 12&20.6 & 20.5&  3.6  &
\\ 
(120178) 2003 OP32& 43.45&0.108&27.1& 41.36&  6&19.9 & 19.9&  4.1  &
\\ 
(120181) 2003 UR292& 32.49&0.176& 2.7& 26.87&  6&21.4 & 21.5&  7.0  &
\\ 
2003 UZ117& 44.29&0.133&27.4& 39.46&  6&22.0 & 21.7&  5.3  &
\\ 
(90377) 2003 VB12&510.00&0.850&11.9& 88.31&  8&21.1 & 21.0&  1.6  &
\\ 
(136204) 2003 WL7& 20.17&0.259&11.2& 15.21&  6&20.6 & 20.6&  8.7  &
\\ 
(175113) 2004 PF115& 39.18&0.062&13.4& 41.34&  6&20.5 & 20.3&  4.7  &
\\ 
2004 PG115& 92.08&0.605&16.3& 36.65&  5&20.8 & 20.9&  5.0  &
\\ 
(120347) 2004 SB60& 42.27&0.105&23.9& 43.89& 10&20.4 & 20.5&  4.2  &
\\ 
2005 CB79& 43.15&0.140&28.7& 40.16&  6&20.4 & 20.3&  5.0  &
\\ 
(145451) 2005 RM43  & 91.37&0.616&28.8& 35.19&  7&19.9 & 19.9&  4.4  &
\\ 
(145452) 2005 RN43 & 41.77&0.028&19.2& 40.72& 13&20.0 & 20.0&  3.9  &
\\ 
(145480) 2005 TB190& 76.58&0.397&26.4& 46.45&  7&20.9 & 20.8&  4.7  &
\\ 
2006 SX368& 22.28&0.463&36.3& 12.44&  4&20.3 & 20.3&  9.5  &
\\ 
2007 JF43& 39.41&0.185&15.1& 39.45&  4&20.9 & 20.8&  5.2  & 3:2
\\ 
2007 JF45& 44.69&0.147&10.6& 38.12&  1d&21.5 & 21.4&  6.0  &
\\ 
2007 JJ43& 48.22&0.166&12.0& 41.96&  3&20.8 & 20.7&  4.9  &
\\ 
2007 JK43 & 46.35&0.492&44.9& 23.93&  3&20.8 & 21.1&  7.6  &
\\ 
2007 NC7& 34.39&0.507& 6.3& 20.37&  3&21.4 & 21.6&  8.6  &
\\ 
2007 OC10& 50.09&0.292&21.7& 35.48&  3&20.8 & 20.8&  5.7  &
\\ 
(225088) 2007 OR10& 67.34&0.500&30.7& 85.37&  7&21.5 & 21.4&  1.9  &
\\ 
2007 PS45& 43.75&0.090&18.9& 39.80&  1d&21.5 & 21.1&  5.6  &
\\ 
2007 RG283& 19.98&0.233&28.8& 18.70&  3&21.5 & 21.0&  8.8  &
\\ 
2007 RH283& 15.96&0.339&21.4& 17.48&  8&21.4 & 21.2&  8.4  &
\\ 
2007 RT15& 39.61&0.234&12.9& 30.90&  3&21.6 & 21.3&  6.9  & 3:2
\\ 
2007 RW10& 30.40&0.303&36.0& 26.24&  7&21.3 & 21.1&  6.5  &
\\ 
(229762) 2007 UK126& 73.52&0.488&23.4& 45.96&  9&20.4 & 20.3&  3.4  &
\\ 
2007 XV50& 46.02&0.073&22.9& 46.19&  3&21.2 & 21.3&  5.0  &
\\ 
2008 AP129& 41.66&0.138&27.4& 37.39&  5&20.6 & 20.7&  5.3  &
\\ 
2008 CS190& 42.08&0.153&16.0& 36.17&  2&21.6 & 21.6&  6.4  & 5:3
\\ 
2008 CT190& 52.47&0.339&38.9& 34.77&  2&21.0 & 21.4&  5.5  & 7:3
\\ 
2008 LP17& 88.04&0.660&14.1& 30.26&  2&21.0 & 20.9&  6.6  &
\\ 
2008 NW4& 45.58&0.203&23.1& 36.92&  2&21.2 & 21.0&  6.0  &
\\ 
2008 OG19& 67.37&0.428&13.1& 38.74&  2&21.6 & 21.3&  4.9  &
\\ 
2008 QB43& 43.36&0.219&26.3& 38.79&  3&21.6 & 21.4&  5.6  &
\\ 
2008 QY40& 63.09&0.418&25.1& 38.11&  2&20.9 & 20.9&  5.3  &
\\ 
2008 SO266& 39.64&0.247&18.8& 31.58&  2&21.5 & 21.4&  6.9  & 3:2
\\ 
2008 SP266& 41.21&0.124&19.5& 36.18&  2&21.2 & 21.2&  5.7  &
\\ 
2008 ST291&106.00&0.607&20.7& 56.68&  2&21.8 & 21.3&  4.4  &

\label{tab:kbos}
\end{longtable}

\begin{table}
\centering
\begin{tabular}{ c c c c c c c  }
\hline
\hline
cluster central  density  & \multicolumn{5}{c}{$\alpha$} \\

\cline{2-6}
 $(M_{\odot}/pc^3)$ & $0.2$ &  $0.35$ &  $0.4$ & $0.6$  & 0.82\\
 \hline
\hline
$10^4$ & 60& 54 & 48 &40 & 47 \\
$10^5$ & 99 & 98 & 99 & 98 &97 &  \\
$10^6$ &100 & 100 &100 & 100 & 100  \\
\hline
\hline
\end{tabular}
 \caption{3D KS test results for the  \cite{2006Icar..184...59B} cluster produced single detections compared to Sedna's orbit. We report the confidence level at which we can reject the two distributions as drawn from the same parent population.}
\label{tab:cluster}
\end{table}

\end{document}